\documentclass[screen,acmsmall]{acmart}
\PassOptionsToPackage{prologue,table,xcdraw}{xcolor}

\AtBeginDocument{%
  \providecommand\BibTeX{{%
    \normalfont B\kern-0.5em{\scshape i\kern-0.25em b}\kern-0.8em\TeX}}}

\setcopyright{acmcopyright}
\copyrightyear{2026}
\acmYear{2026}
\acmDOI{XXXXXXX.XXXXXXX}

\graphicspath{{./figs/}}
\DeclareGraphicsExtensions{.pdf,.jpeg,.png,.eps}

\usepackage{colortbl}
\usepackage{caption}
\usepackage{subcaption}
\usepackage{booktabs, tabularx}
\usepackage{threeparttable}
\usepackage{longtable}

\usepackage{algorithm}
\usepackage{algpseudocode}
\floatname{algorithm}{Procedure}
\usepackage{enumerate}
\usepackage{fancybox}  
\usepackage{multirow}
\usepackage{pifont}
\usepackage{xspace}

\usepackage{listings}
\usepackage{makecell}
\usepackage{float}
\usepackage[export]{adjustbox}
\usepackage{enumitem}

\newcommand{\secref}[1]{Section~\ref{#1}}
\newcommand{\tabref}[1]{Table~\ref{#1}}

\DeclareTextFontCommand{\emp}{\bfseries}

\usepackage[most]{tcolorbox}
\definecolor{custom-gray}{cmyk}{0, 0, 0, 0.7, 1.00}
\newtcolorbox{Summary}[2][]{
top=0.15in,
fonttitle=\bfseries,
colbacktitle=custom-gray,
colback=gray!5,
colframe=gray!40!black,
enhanced,
attach boxed title to top left={xshift=1.5em,yshift=-\tcboxedtitleheight/2},
boxed title style={size=small,colback=custom-gray},
drop shadow={black!50!white},
title=#2,#1}

\newcommand{\model}[1]{\textsc{#1}}
\newcommand{\rqtwoverb}{augment\xspace}
\newcommand{\rqtwonoun}{augmentation\xspace}
\newcommand{\rqtwoaction}{augmenting\xspace}
\newcommand{\rqtwoactionpast}{augmented\xspace}
\newcommand{\rqtwoactor}{augmentor\xspace}

\definecolor{mygreen}{RGB}{0,158,115}
\definecolor{myyellow}{RGB}{230,159,0}
\definecolor{myorange}{RGB}{255, 87, 37}
\definecolor{mycyan}{RGB}{0, 173, 181}
\definecolor{myblue}{RGB}{0, 114, 178}
\newcommand{\gopifix}[1]{}
\newcommand{\haofix}[1]{}

\begin{document}
\graphicspath{{figs/}}

\title[Towards Improving AI Agent Efficiency with Augmented MCP Tool Descriptions]{Model Context Protocol (MCP) Tool Descriptions Are Smelly! Towards Improving AI Agent Efficiency with Augmented MCP Tool Descriptions}

\author{Mohammed Mehedi Hasan}
\orcid{0000-0001-9837-0998}
\email{mohammedmehedi.hasan@queensu.ca}
\affiliation{%
  \institution{Queen's University}
  \city{Kingston}
  \state{ON}
  \country{Canada}
}
\author{Hao Li}
\orcid{0000-0003-4468-5972}
\affiliation{%
  \institution{Queen's University}
  \city{Kingston}
  \state{ON}
  \country{Canada}
}
\email{hao.li@queensu.ca}

\author{Gopi Krishnan Rajbahadur}
\orcid{0000-0003-1812-5365}
\affiliation{%
  \institution{Queen's University}
  \department{School of Computing}
  \city{Kingston}
  \state{Ontario}
  \country{Canada}
}
\email{grajbahadur@acm.org}

\author{Bram Adams}
\orcid{0000-0001-7213-4006}
\affiliation{%
  \institution{Queen's University}
  \city{Kingston}
  \state{ON}
  \country{Canada}
}
\email{bram.adams@queensu.ca}

\author{Ahmed E. Hassan}
\orcid{0000-0001-7749-5513}
\affiliation{%
  \institution{Queen's University}
  \city{Kingston}
  \state{ON}
  \country{Canada}
}
\email{ahmed@cs.queensu.ca}

\renewcommand{\shortauthors}{M. Mehedi Hasan et al.}

\begin{abstract}
The Model Context Protocol (MCP) introduces a standard specification that defines how Foundation Model (FM)-based agents should interact with external systems by invoking tools. However, to understand a tool's purpose and features, FMs rely on natural-language tool descriptions, making these descriptions a critical component in guiding FMs to select the optimal tool for a given (sub)task and to pass the right arguments to the tool. While defects or smells in these descriptions can misguide FM-based agents, their prevalence and consequences in the MCP ecosystem remain unclear. 

Hence, we examine 856 tools spread across 103 MCP servers empirically, assess their description quality, and their impact on agent performance. We identify six components of tool descriptions from the literature, develop a scoring rubric utilizing these components, and then formalize tool description smells based on this rubric. By operationalizing this rubric through an FM-based scanner, we find that 97.1\% of the analyzed tool descriptions contain at least one smell, with 56\% failing to state their purpose clearly. While \rqtwoaction these descriptions for all components improves task success rates by a median of 5.85 percentage points and improves partial goal completion by 15.12\%, it also increases the number of execution steps by 67.46\% and regresses performance in 16.67\% of cases. These results indicate that achieving performance gains is not straightforward; while execution cost can act as a trade-off, execution context can also impact. Furthermore, component ablations show that compact variants of different component combinations often preserve behavioral reliability while reducing unnecessary token overhead, enabling more efficient use of the FM context window and lower execution costs.
\end{abstract}




\maketitle
\newcommand{\rqzero}{How healthy and sustainable are MCP servers?}
\newcommand{\rqone}{To what extent do MCP servers contain security vulnerabilities?}
\newcommand{\rqtwo}{To what extent do MCP servers contain maintainability issues?}

\newcommand{\motivation}{\emp{Motivation. }}
\newcommand{\approach}{\medskip\noindent\emp{Approach. }}
\newcommand{\findings}{\medskip\noindent\emp{Findings. }}
\newcommand{\runningexample}{\textbf{Running example. }}


\section{Introduction}\label{sec:introduction}
Model Context Protocol (MCP) adoption for Foundation Models (FMs), such as GPT, continues to grow across domains, including health~\cite{ehtesham2025enhancing}, bioinformatics~\cite{widjaja2025bioinfomcp}, transportation~\cite{chhetri2025model}, vision systems~\cite{tiwari2025model}, and especially software engineering~\cite{sarkar2025survey}. MCP provides a unified interface to bridge FM-based agents with external capabilities (tools) by exposing them to the following three tools-related natural-language artifacts: a tool name, a tool description, and an input schema (with argument names and their data types) to FMs. This purely native natural-language-based alignment of MCP with the agentic ecosystem is driving its massive adoption, leading several major companies, including GitHub, Google Cloud Platform (GCP), and PayPal, to develop and maintain their own MCP servers, commonly referred to as \textit{official MCP servers}~\cite{hasan2025model}. In parallel, independent developers and open-source contributors have created numerous \textit{community MCP servers} that integrate a wide range of third-party services~\cite{hasan2025model}. 

\begin{figure}[t]
    \centering
    \includegraphics[width=.95\textwidth]{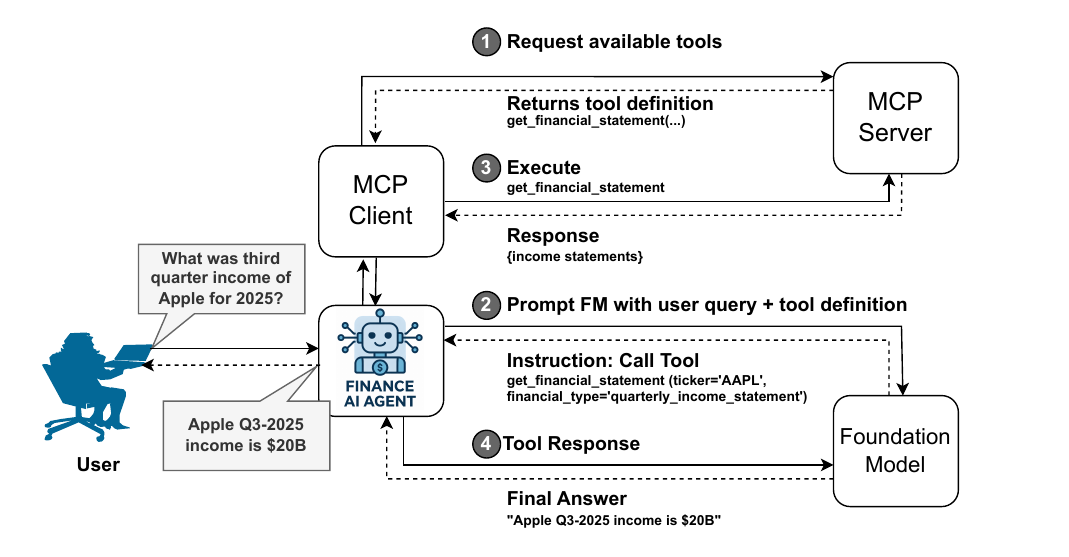}
    \caption{MCP workflow for an FM-based agent. When an agent receives a user query, (1) it retrieves tool metadata (name, description, and input schema) via the MCP client; (2) the agent prompts the foundation model (FM) with the user query and retrieved metadata, whereupon the FM plans the solution, formulates the appropriate tool call, and instructs the agent to execute it; (3) the agent executes the tool call via the MCP client; and (4) the agent forwards the tool response to the FM, which synthesizes the final answer for the user.}
    \label{fig:intro:toolcall}
\end{figure}


In MCP-enabled workflows with these official and community MCP servers, tool descriptions serve as the essential linguistic guide that directs FM behavior. These descriptions convey a tool's intended functionality, constraints, and usage cues, and shape tool selection, parameterization, and multi-step orchestration~\cite{hou2025model}. For example, as illustrated in Figure~\ref{fig:intro:toolcall}, upon receiving a user query (e.g., ``\textit{What was the third-quarter income of Apple for 2025?}''), the agent retrieves the names, descriptions, and schemas of all available tools from the connected MCP servers and injects this metadata into the FM’s context, along with the user query. Only then can the FM leverage these artifacts to discover the capabilities of the available tools, plan a solution strategy, select an appropriate tool (\texttt{get\_financial\_statement}), infer the required parameters  (\texttt{ticker="AAPL", financial\_type="quarterly\_income\_statement"}), and issue a tool call via the agent's MCP client. The agent then returns the tool response to the FM, which synthesizes the final answer for the user.

From this flow, it is quite clear that if the tool descriptions are defective, underspecified, or misleading, the FM may select the wrong tool, supply invalid or suboptimal arguments, or take unnecessary interaction steps, ultimately reducing the reliability of MCP-enabled systems. In other words, the tool description is not merely documentation, but embodies a dual nature, as it serves as (i) a requirement-like specification that defines the tool's expected behavior and parameter constraints~\cite{stoica2024specifications}, and (ii) a prompt-like instruction that shapes the model's contextual reasoning and decision-making~\cite{mei2025survey}. This hybrid role blurs the boundary between software requirements and natural-language prompts, creating a novel design surface where textual or structural imperfections can propagate in the form of specification errors and prompt misguidance. We conceptualize these imperfections as tool description smells, similar to the concept of recurring suboptimal patterns that degrade clarity, correctness, or maintainability~\cite{moha2009decor,vogelsang2025impact} in software engineering. 

While smells in MCP code have been reported~\cite{hasan2025model}, the prevalence and distribution of smells in MCP tool descriptions remain largely unexplored. Prior work on prompts has shown that instructional artifacts are composed of multiple components, including personality information, task information covering task intent, user demand, and domain information, as well as demonstration through examples, which together contribute to the accuracy and efficiency of FM~\cite{liu2026comprehensive}. We suggest that MCP tool descriptions exhibit an analogous component structure and that smells can arise at the component level.

On the other hand, the efficiency of AI agents using MCP servers has recently come under strict scrutiny, as the tool metadata, repeatedly injected into the FM's context during a typical interaction with an FM-based agent, is inflating token usage and increasing execution cost. While emerging techniques like \texttt{Agent Skills}~\cite{agent-skills} attempt to complement the MCP's use by progressive discovery of capabilities to an FM, and \texttt{Tool Search}~\cite{tool-search} is enabling the FM to search for a tool on demand rather than always loading them into context, they do not eliminate the reliance on MCP tools and their tool descriptions. Instead, these developments highlight a fundamental tension in MCP-enabled agents: resolving tool description smells by \rqtwoaction all components may improve semantic guidance and agent performance, but it also consumes scarce context windows and increases costs. Any attempt to \rqtwoverb tool descriptions must therefore justify its cost and, ideally, identify compact representations that preserve effectiveness.

Despite extensive work on FM tool calling challenges, including complex function calls~\cite{zhong2025complexfuncbench}, large tool sets~\cite{qin2023toolllm}, and security attacks on tool selection~\cite{shi2025prompt, wang2025mpma}, there has been no systematic investigation of the quality of tool descriptions in MCP servers or their downstream impact on agent performance. Although industry documentation and practitioner guidelines propose best practices for writing tool descriptions~\cite{anthropic-tool-description, xu2025llm, asturn-mcp-tools}, it remains unclear how widely such practices are adopted in the MCP ecosystem and whether they really improve agent behavior under realistic workloads. 

Recognizing this lack of systematic research, we carry out an extensive empirical study of MCP tool description quality and its effect on FM-based agent performance. On a dataset of 103 major MCP servers comprising 856 tools, we scan tool descriptions using a structured quality rubric designed to identify potential suboptimal design patterns or smells~\cite{moha2009decor}. We then use FMs to automatically fix the identified smells and \rqtwoverb the description. Finally, we utilize the MCP Universe benchmark~\cite{luo2025mcp} to evaluate how these \rqtwoactionpast descriptions influence FM-based agents' performance. This evaluation is structured around the following core Research Questions (RQs).

\medskip\noindent\textbf{RQ-1: To what extent do MCP tools’ descriptions contain smells?}

\motivation
Despite the critical role of tool descriptions as both requirement-like specifications and prompt-like instructions, their quality in real-world MCP deployments remains largely unexamined. In particular, there is no empirical baseline characterizing which components tool descriptions typically include in practice, how frequently they exhibit smells, or how these smells differ between official and community-maintained MCP servers. Prior research shows that smells in software artifacts (e.g., code, tests, datasets, and prompts) increase change-proneness and erode reliability~\cite{khomh2009exploratory, hassan2022code, zhao2025workflows, ronanki2024prompt}, suggesting similar risks for MCP-enabled agents. We therefore quantify the prevalence and distribution of tool description smells across both official and community-maintained MCP servers.

\emp{Findings.}
We find that 56\% of the 856 MCP tool descriptions exhibit an \textit{Unclear Purpose} smell, indicating that a majority fail to articulate their intended functionality clearly to the FM. More broadly, 97.1\% of tool descriptions contain at least one smell, and the majority exhibit multiple smell types, particularly Unstated Limitations, Missing Usage Guidelines, and Opaque Parameters affecting them. Tool descriptions from both official and community-maintained servers exhibit these issues, indicating that producing high-quality tool descriptions is challenging for all types of practitioners. 
 
\medskip\noindent\textbf{RQ-2: How does resolving tool description smells by \rqtwoaction all tool description components impact the performance of FM-based agents?}

\motivation
Given that RQ-1 reveals that tool description smells are widespread, a natural scientific question is whether resolving them by \rqtwoaction  underspecified or missing components improves agent behavior in realistic MCP workflows. Prior software engineering research reports mixed effects from smell removal. While it improves specific quality attributes, such as energy efficiency and runtime performance in some contexts~\cite{cedrim2017understanding}, it can also unintentionally alter system behavior in others~\cite{verdecchia2018empirical}. In contrast, prompt enhancement techniques for FMs, such as DSPy~\cite{khattab2023dspy}, MIPROv2~\cite{opsahl2024optimizing}, and GEPA~\cite{agrawal2025gepa}, have consistently demonstrated performance gains. Motivated by these conflicting signals, we investigate whether \rqtwoaction MCP tool descriptions with all components improves agent performance in practice and whether such improvements entail trade-offs, if any.

\emp{Findings.}
Augmented tool descriptions yield a statistically significant increase of 5.85 percentage points in task success rate across domain-model combinations, while causing regressions in 16.67\% of cases in the MCP Universe benchmark. They also improve evaluator-level performance, increasing the Average Evaluator score by 15.12\%, reflecting higher-quality intermediate execution step completion. These improvements come with a trade-off: the average number of execution steps increases by 67.46\% (median), indicating that agents expend significantly more interaction steps with richer descriptions. Analysis of this accuracy-cost trade-off reveals that different domain-model combinations support distinct operating points, allowing practitioners to prioritize either peak accuracy or lower execution costs, depending on their deployment requirements. 

\medskip\noindent\textbf{RQ-3: How do different components of the \rqtwoactionpast tool description impact the performance of FM-based agents?}

\motivation
The results of RQ-2 demonstrate that fully \rqtwoactionpast tool descriptions improve agent performance but incur substantial execution overhead, intensifying the tension between the semantic completeness and token efficiency. Practitioners warn that excessive detail can saturate the FM’s context window,\footnote{\url{https://jentic.com/blog/the-mcp-tool-trap}} making fully \rqtwoactionpast descriptions impractical for use cases where the context window is scarce. Furthermore, prior research indicates that not all components of instructional prompts contribute equally to FM behavior~\cite{yin2023did}. We therefore investigate the impact of individual tool description components through an ablation study to identify a minimal effective set that preserves performance while reducing context overhead.

\emp{Findings.}
Our results indicate that no single combination of MCP tool description components consistently yields improved performance across all domains and models. However, shorter, targeted descriptions can retain the core semantic content of fully \rqtwoactionpast descriptions while achieving statistically equivalent performance. Across all domain-model combinations, removing the \textit{Examples} component does not statistically degrade performance. These findings indicate that practitioners can identify the most impactful components for their specific domain and model as a lower-cost alternative without sacrificing effectiveness. 


\vspace{1em}

Our study makes the following key contributions to MCP ecosystem:

\begin{enumerate}
    \item \textbf{Scoring Rubric}: We consolidate best practices for writing MCP tool descriptions from multiple sources and propose the first structured scoring rubric to evaluate the quality of individual description components.
    \item \textbf{FM-based Smell Scanner}: Using this rubric, we develop the first automated smell detector for MCP tool descriptions, released as part of our replication package~\footnote{\url{https://github.com/SAILResearch/mcp-tool-description-augmentation}}, enabling developers to identify quality issues in their tools.
    \item \textbf{Tool Description Augmentor}: We introduce an FM-based \rqtwoactor that systematically resolves smells by enriching tool descriptions with all the components that practitioners can leverage to fix smells.
    \item \textbf{Tool Description Router}: We present the first tool description router that allows MCP users to experiment with multiple versions of a tool description at runtime and select the variant that performs best in their workflow without changing the code of the MCP servers.
    \item \textbf{Empirical Findings}: Through large-scale analysis and benchmark evaluation, we quantify smell prevalence, demonstrate performance effects of \rqtwonoun, and identify the impact of different description components on the performance. 
\end{enumerate}

\section{A Motivational Example}\label{sec:motivation}
\begin{figure}[t]
    \centering
    \begin{subfigure}[b]{\textwidth}
        \centering
        \includegraphics[width=\linewidth, frame]{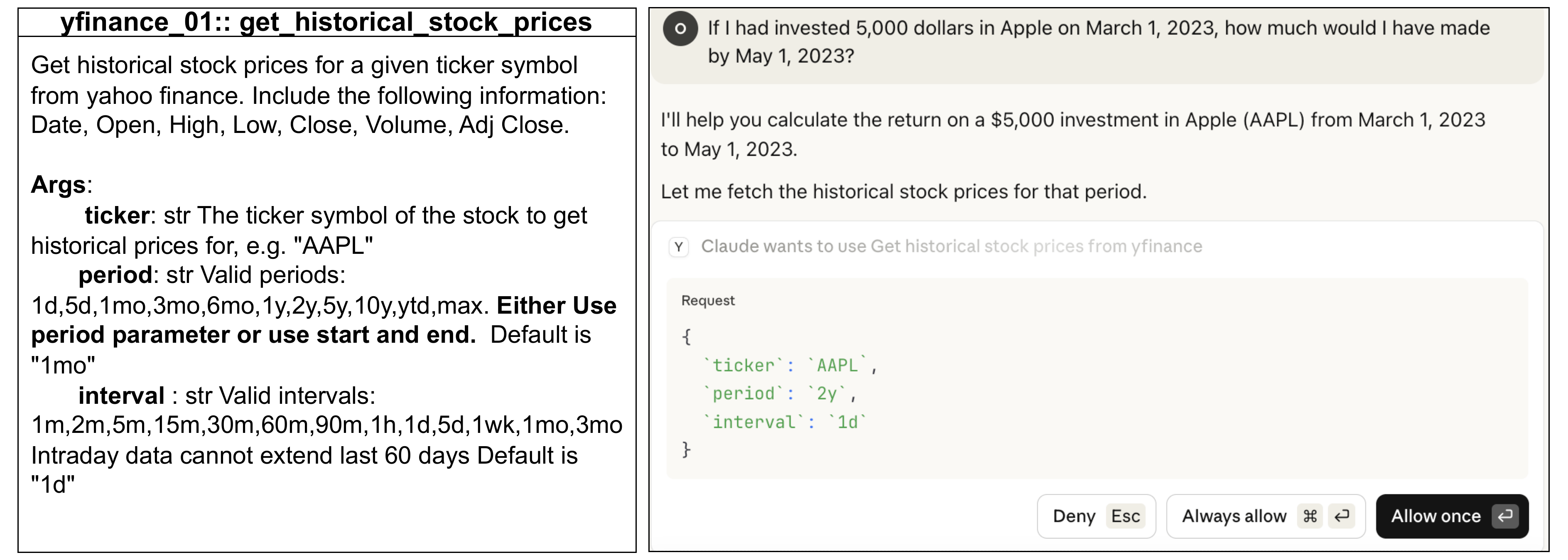}
        \caption{Original Yahoo Finance MCP tool description for \texttt{get\_historical\_stock\_prices} showing only reference to “start” and “end” without explicit argument names or format specification.}
        \label{fig:motivating-example-yf-01}
    \end{subfigure}
    \hfill
    \begin{subfigure}[b]{\textwidth}
        \centering
        \includegraphics[width=\linewidth, frame]{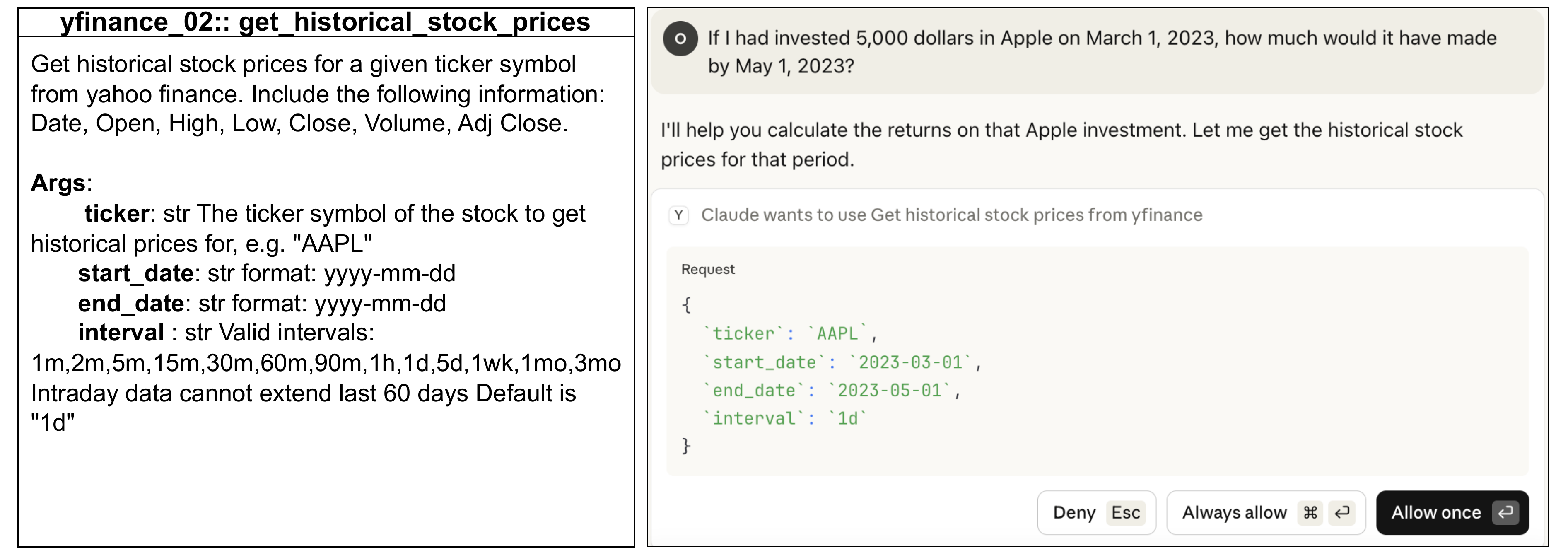}
        \caption{Forked Yahoo Finance MCP tool description defining explicit arguments \texttt{start\_date} and \texttt{end\_date} with specified \texttt{yyyy-mm-dd} format.}
        \label{fig:motivating-example-yf-02}
    \end{subfigure}
    \caption{Comparison of two Yahoo Finance MCP tool descriptions used by the same FM-based agent. The original version (a) provides ambiguous guidance, while the forked version (b) clarifies parameter names and formats. This difference in description quality directly influences how the FM selects parameters during tool invocation, affecting data retrieval scope, latency, and overall efficiency.}
    \label{fig:motivating-example}
\end{figure}

FM-based agents rely on MCP servers to reach external capabilities. As seen in Figure~\ref{fig:motivating-example-yf-01}, how an agent plans and invokes those tools is shaped by the tool descriptions the FM reads. We illustrate a common failure seen in real deployments and how a small change in description shifts agent behavior.

Consider a scenario where Alex is an AI engineer at a financial institution tasked with building a finance assistant that answers portfolio questions and simple what-if queries. The team selects a popular Yahoo Finance MCP server and wires it to a frontier FM.

\textbf{Phase 1: It works.} During the development phase, the agent handles simple requests, for example, ``show the last month of prices'' and ``plot recent history.'' The agent: (1) forwards the user request and tool list to the FM, (2) the FM picks the historical-price tool, (3) the tool returns data, (4) the agent responds cleanly. The team ships the agent.

\textbf{Phase 2: The unseen inefficiency.} Days later, users start asking time-bounded questions, for example, ``What happened around last March?'' Responses slow down, costs start to rise, and logs show large payloads. Traces reveal that the FM is calling \texttt{get\_historical\_stock\_prices} with a broad \texttt{period} that expands to multi-year windows, despite the users' question regarding narrow time periods. This inflates the response size of the tool and downstream token usage of the FM.

\textbf{Phase 3: Root cause in the description.} The original tool description (Fig.~\ref{fig:motivating-example-yf-01}) lists a \texttt{period} parameter and says “Either use period parameter or use start and end,” but never names \texttt{start} or \texttt{end} as explicit parameters, nor gives the data type or format (e.g., whether yyyy-mm-dd or dd-mm-yyyy). Lacking concrete parameter names and guidance, the FM cannot reliably infer how to construct a bounded time range and therefore defaults to the \texttt{period} parameter, often selecting ranges broader than required for the user query. This is not a model bug; it is a specification problem in the tool description.

\textbf{Phase 4: A small fix that changes behavior.} The team tries a forked MCP server whose tool description replaces the vague mentions of \texttt{start\_date} and \texttt{end\_date} with explicit arguments, specifying the expected data format \texttt{yyyy-mm-dd} (Figure~\ref{fig:motivating-example-yf-02}). With clear names and constraints, the FM starts issuing date-bounded calls that fetch only the needed window, reducing upstream data volume, latency, and token cost, while improving answer relevance for time-scoped queries.

\textbf{Phase 5: The challenge.} This experience shows that seemingly minor description details can materially affect agent performance, prompting Alex and their team to reflect on the existence of a broader set of unanswered questions that should be considered when integrating third-party MCP servers into production agents. In particular, teams deploying FM-based agents must understand:

\begin{enumerate}
    \item How prevalent quality issues or smells are in MCP tool descriptions in practice, and how these issues are distributed across official and community-maintained MCP servers (\textbf{RQ1}).
    \item Whether systematically \rqtwoaction underspecified or missing tool description components improves FM-based agent performance in realistic MCP workflows, and what trade-offs such improvements entail (\textbf{RQ2}).
    \item What components of a tool description most strongly impact the performance of FM-based agents, and whether there exists any generalizable ``golden rule'' for deriving effective tool descriptions across different MCP servers and use cases (\textbf{RQ3}).
\end{enumerate}
\section{Background and Related Work}\label{sec:background}
\subsection{Model Context Protocol (MCP)}
To let AI applications operate on real systems, FM-based agents rely on tools that perform external actions, e.g., searching the web for information retrieval, executing database queries, triggering API calls, running programs, or performing device management. Anthropic proposed the Model Context Protocol (MCP) so that AI agents can discover tools using a standard client-server protocol and invoke them with the proper arguments, reducing the need for custom glue code across models and frameworks. Since then, MCP has been evolving as an open integration protocol that facilitates communication between AI applications and external systems, often positioned as a ``USB-C port for AI''~\cite{anthropic_mcp}. Since its introduction, the protocol has been adopted by major FM providers such as OpenAI, Microsoft, Google, and Cloudflare~\cite{hasan2025model}, and currently observes over 20 million weekly downloads for the Python and JavaScript SDKs of MCP servers, signaling strong community acceptance and usage.

To enable this interoperability at scale, MCP adopts a client-server architecture that cleanly separates AI agents from tool implementations. In this design, an AI agent spins up one MCP client to connect to MCP servers. Each server can run locally or remotely and may expose multiple tools, resources, and prompts~\cite{hou2025model}. Discovery and capability negotiation follow a JSON-RPC data layer with an initialization handshake and list/get methods that let clients enumerate available tools before execution through a well-known protocol called reflection~\cite{hasan2025model}. MCP supports multiple transport options including stdio for local, process-to-process connections, and streamable HTTP with optional server-sent events for remote servers and authenticated access.

Within this architecture, the primary handshaking interface between MCP servers and foundation models is the tool description itself. Each MCP server exposes its tools through a structured description consisting of a name, a natural-language description, and an input schema. Through reflection, this information is passed to the Foundation Model (FM) by MCP clients, which rely on it to select and invoke the correct tool. As shown in Figure~\ref{fig:motivating-example}, each tool in an MCP server should clearly describe its purpose (i.e., the core functionality it provides) and guide the FM on how to use it. For example, the description of the \texttt{get\_financial\_statement} tool specifies what it does, retrieving financial statements for a company from Yahoo Finance, and how to use it by mentioning the types of statements that can be obtained. A well-written tool description therefore conveys the tool's purpose, usage guidance, and any relevant caveats or examples~\cite{anthropic-tool-description, qu2025tool}.

These descriptions play a central role at runtime, shaping how agents reason about and execute tool calls. As shown in Figure~\ref{fig:intro:toolcall}, MCP standardizes how FM-based agents interact with tools through a client-server loop. The MCP client orchestrates this communication loop among the FM and available MCP servers, each of which exposes tool metadata, e.g., name, description, and input schema, to the model.
\begin{enumerate}
    \item \textbf{Discovery} The client queries connected servers via reflection to list available tools and their metadata. In the finance example, it retrieves entries such as get\_financial\_statement or get\_historical\_stock\_prices from the yahoo-finance-mcp-server.
    \item \textbf{Planning} The client embeds these descriptions in the FM’s context along with the user query. Using its language reasoning, the FM selects the correct tool and infers the tool's arguments, e.g., \texttt{ticker="AAPL" and financial\_type="quarterly\_income\_statement"}.
    \item \textbf{Execution} The FM issues a tool-call instruction. For this, the MCP client validates parameters, seeks user consent for sensitive actions, and executes the call through the appropriate server. The FM then synthesizes the final answer from the returned data.
    \item \textbf{Reuse} Because all interaction occurs through the MCP interface, the same server (e.g., yahoo-finance-mcp-server) can be reused across different agents and frameworks without re-implementation.
\end{enumerate}


Given this neatly coupled reasoning loop between the FM and tool descriptions, evaluating MCP-enabled agents requires benchmarks that can faithfully capture both planning correctness and execution behavior. In this study, we adopt the MCP-Universe benchmark~\cite{luo2025mcp}, which is one of the most comprehensive and widely used benchmarks for evaluating MCP-based agents. MCP-Universe spans multiple domains, such as finance, data analysis, repository management, and information retrieval, and defines realistic, goal-oriented tasks that test an agent's ability to choose the correct tool and execute it with the proper arguments. It provides a total of 18 MCP servers with 202 tools combined. Each task is evaluated by at least one evaluator, with an average of 3.3 evaluators per task.
\subsection{Studies on Smells}
Software “smells” have long been studied as indicators of latent design or process issues rather than explicit faults. Fowler and Beck characterize smells as weaknesses that may slow development or increase future error risk without being technically incorrect~\cite{fowler2018refactoring}. Because this notion links smells to long-term maintenance costs, the topic has received sustained attention.

Early research studied \emph{code smells} through taxonomies and catalogs~\cite{mantyla2003taxonomy, marticorena2006extending, jerzyk2023code}, followed by empirical work that examined their evolution and impact~\cite{olbrich2009evolution, yamashita2012code, sjoberg2012quantifying}. These studies recommend practices such as limiting module size and avoiding large multi-purpose changes to improve maintainability. Detection techniques span textual heuristics~\cite{vislavski2018licca}, repository mining~\cite{palomba2014mining}, and token-based analysis~\cite{kamiya2002ccfinder, wang2018ccaligner} across multiple languages.

Smells are not limited to source code. \emph{Architectural smells}, such as \textit{connector envy} and \textit{ambiguous interface}, have been identified~\cite{garcia2009toward, garcia2009identifying}, with tools like \textit{Arcan} supporting automated detection~\cite{fontana2017arcan}. \emph{Test smells}, e.g., \textit{assertion roulette}, \textit{mystery guest}, and \textit{eager test}, are also prevalent and can hinder comprehension and maintenance~\cite{bavota2012empirical, bavota2015test, tufano2016empirical}. Beyond implementation artifacts, requirements and design artifacts exhibit smell-like deficiencies~\cite{femmer2017rapid, khomh2009bayesian, moha2009decor}, and defects traced to later phases are known to be substantially more costly to remedy.

With the advent of foundation models (FMs), newer categories of smells have emerged. These include smells in FM-generated code~\cite{siddiq2024quality, paul2025investigating} and unit test code~\cite{ouedraogo2024test}, as well as data-related smells, such as data leakage, lack of context, or misleading instances, observed in FM-driven systems~\cite{vitale2025catalog}. In addition, \emph{prompt smells}~\cite{ronanki2024prompt} have been identified as factors that degrade FM output quality by introducing ambiguity, bias, or inconsistency in instruction formulations.

Within the MCP ecosystem, recent work reports code smells in MCP servers~\cite{hasan2025model}. However, to date, no empirical analyses have examined their design- or specification-level ``smells``. In this study, we focus on tool descriptions as a first-class artifacts and, by analyzing the smells hidden within them, we provide empirical evidence on the impact of tool descriptions on agent performance.

\subsection{Refactoring the Smells and Optimization}
In classical software engineering research, automated code smell identification and refactoring have primarily relied on static analysis techniques grounded in heuristics, coupling-cohesion metrics, and distance-based similarity measures to detect smell manifestations and suggest behavior-preserving refactorings~\cite{tsantalis2009identification,tsantalis2011identification}. Building on these foundations, more recent refactoring surveys have systematically categorized these efforts into broader families of approaches, including metrics- and precondition-oriented methods, clustering- and graph-based analyses, code slicing and dynamic analysis techniques, as well as search-based optimization strategies~\cite{lacerda2020code}.


Beyond static analysis, a growing body of work has operationalized different machine learning (ML) and deep learning (DL) based techniques to identify refactoring opportunities and predict refactoring actions. These include multi-layer perceptrons (MLPs) and recurrent neural network (RNN) variants such as bidirectional long short-term memory (BiLSTM) and gated recurrent unit networks (GRU) for refactoring type prediction~\cite{mohan2016technical,szalontai2021detecting}, convolutional neural network (CNN)- and RNN-based models with embedding pipelines for naming-related refactorings such as Rename Method~\cite{liang2021deep}, and RNN encoder-decoder architectures that model refactoring as a learned code transformation task~\cite{tufano2019learning}, as summarized in the survey by Naik et al.~\cite{naik2024deep}.

Recent advances in FMs have expanded this space by introducing FM-based strategies for automated smell detection and correction. For instance, \textit{iSmell}~\cite{wu2024ismell} integrates multiple smell detection toolsets through a Mixture of Experts (MoE) architecture to identify and refactor code segments with smelly code. Similarly, techniques such as Co-pilot loops~\cite{zhang2024copilot} employ agentic feedback cycles to iteratively refine code, while \textit{UTRefactor}~\cite{gao2025automated} targets smell remediation within unit tests. Beyond code and test refactoring, multi-agent frameworks have also been proposed to address architectural and design-level smells~\cite{pandini2025exploratory}.

Parallel research has explored optimization in the context of prompt engineering. In addition to heuristic-based search methods~\cite{cui2025automatic}, recent studies propose FM-based prompt optimization frameworks such as MAP~\cite{chen2023mapo}, DSPy~\cite{khattab2023dspy}, EASE~\cite{wu2024prompt}, MIPROv2~\cite{opsahl2024optimizing}, and GEPA~\cite{agrawal2025gepa}. These approaches leverage differentiable optimization or feedback-guided refinement and, in several cases (e.g., GEPA), outperform reinforcement learning-based techniques in achieving higher-quality model responses.

Despite the functional similarities between prompts and tool descriptions in the MCP ecosystem, no prior study has investigated the optimization or refactoring of tool descriptions. To fill this gap, we examine how FM-based \rqtwonoun can be adapted to improve the quality of MCP tool descriptions.

\subsection{Evaluating MCP-Enabled AI Agents}
Evaluation of agents has progressed rapidly, yet most popular benchmarks emphasize general agentic or language capabilities rather than the specific competence of utilizing MCP tools effectively. To broaden coverage, recent studies have begun to assess orchestration and tool-use behaviors in realistic settings. For example, \textit{MCP-Universe} spans six domains with 231 tasks, using fine-grained evaluators to measure goal-directed tool sequencing~\cite{luo2025mcp}. \textit{LiveMCPBench} provides multi-domain, multi-server assessment of multi-step trajectories~\cite{mo2025livemcpbench}. Additional efforts include \textit{LIVEMCP-101}, which stress-tests long-horizon queries~\cite{yin2025livemcp}, \textit{MCPWorld} for computer-use agents~\cite{yan2025mcpworld}, \textit{MCPEval} for standard metrics and automated pipelines~\cite{liu2507mcpeval}, and \textit{MCPToolBench++} for multi-domain and multilingual evaluation~\cite{fan2025mcptoolbench++}.

Despite this progress, existing benchmarks maintain static MCP server specifications, i.e., tool names, descriptions, and parameters remain unchanged. Consequently, there is no empirical evidence on whether \rqtwoaction or changing the tool descriptions improves or degrades agent outcomes. Our study addresses this gap by evaluating agents under \rqtwoactionpast tool descriptions and by conducting ablation studies to isolate the impact of individual description components on downstream performance.

\section{Methodology}\label{sec:exp_design}
We illustrate the methodology used in this study in Figure~\ref{fig:methodology} and describe each step in the following sub-sections:

\begin{figure*}[t]
	\centering
	\includegraphics[width=0.99\textwidth]{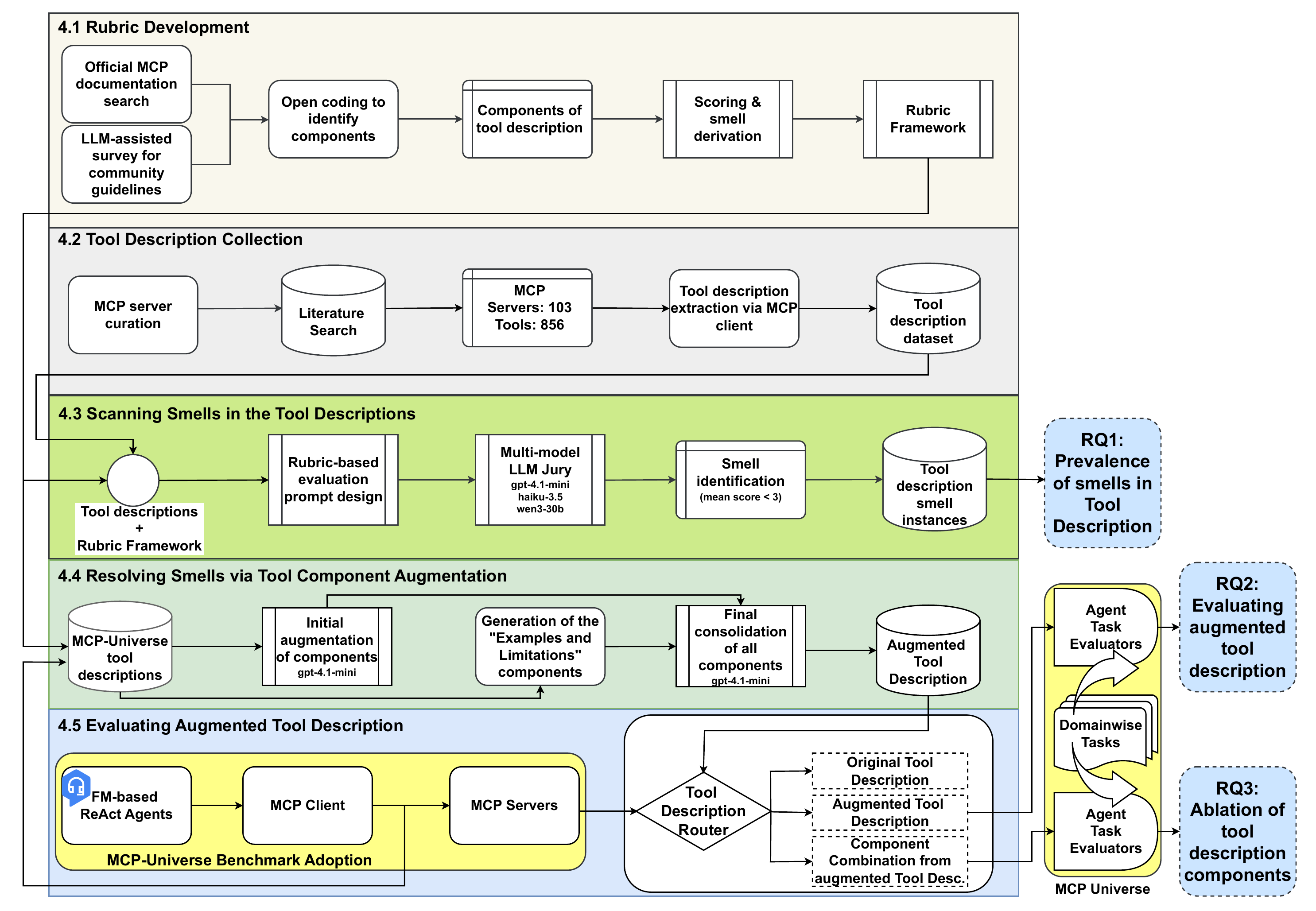}
	\caption{Overview of the process used to study MCP server tool descriptions. The components in bright yellow boxes represent workflows repurposed from the MCP-Universe benchmark for evaluation and benchmarking, while all other components and processes were introduced in this study.}
	\label{fig:methodology}
\end{figure*}

\subsection{Rubric Development}\label{rubric-development}

To systematically evaluate tool descriptions, we move beyond simple qualitative guidelines and construct a structured analytic rubric. Our rubric is a multi-dimensional analytic evaluation framework composed of six components of tool description. Each component is scored independently on a 5-point Likert scale, defined by specific performance descriptors, where Score 3 represents the minimum viable threshold. This rubric development process involves four steps, which we describe below.

\subsubsection{Official MCP documentation search.}
As MCP was originally proposed by Anthropic and their technical documents define the baseline tool standard, we first consult Anthropic’s official MCP documentation~\cite{anthropic-tool-description} when studying the tool descriptions. This documentation provides guidance on the required components of a tool description and outlines how these components support model comprehension.


\subsubsection{LLM-assisted survey for community guidelines}
To capture a broader community consensus beyond the official specification, we conduct an LLM-assisted survey of community guidelines. For this survey, we utilize an LLM-based agentic deep research technique, which is known to systematically survey and synthesize distributed practitioner knowledge more efficiently than traditional keyword-based searches~\cite{zhang2025web, xi2025survey}. We adopt OpenAI Deep Research\footnote{\url{https://openai.com/index/introducing-deep-research/}} with the GPT-5.1 model to supplement the official recommendations. This deep research process identifies 15 sources that address best practices for tool description design, including four tutorials, four blog posts, three Reddit discussions, three research articles~\cite{xu2025llm, qin2023toolllm, hsieh2023tool}, and one GitHub discussion. We provide the prompts and interaction with the Deep Research in our replication package~\footnote{\url{https://github.com/SAILResearch/mcp-tool-description-augmentation}}.

\subsubsection{Open coding to identify components of tool description}\label{deriving-key-component}
We manually analyze all collected sources and synthesize the recurring recommendations to construct a comprehensive list of tool description components. The first two authors independently analyzed the 15 sources and uncovered the components through a small, open coding exercise. Because each source can discuss multiple components, this coding exercise constitutes a multi-label annotation task. We evaluate inter-rater agreement by computing Jaccard similarity for each source and averaging across sources, following prior multi-label annotation studies~\cite{parker2024large, hasan2025empirical}. The mean Jaccard similarity is 0.92, indicating excellent agreement. The resulting six components are grounded in official guidance from Anthropic~\cite{anthropic-tool-description}, practitioner insights~\cite{asturn-mcp-tools, merge-dev-mcp-tools}, and academic analyses~\cite{xu2025llm}. 

We observed how these tool description components each play one of two complementary roles, consistent with the dual nature discussed in~\secref{sec:introduction}. One group of components are requirement-like specification components that describe what the tool does and how it should be invoked, including its purpose, limitations, parameter explanation, and examples. These components define the strict functional contract and constraints required for valid execution. The second group of tool description components represent prompt-like instructional components, such as Guidelines and Length and Completeness, that do not introduce new functional constraints but instead serve as behavioral directives that shape how the FM interprets the description, prioritizes information, and reasons about tool selection and invocation. Below, we define each of the six components and provide an illustrative example drawn from the \textit{Sequential Thinking} tool description shown in Figure~\ref{fig:sequential-thinking-tool-description}:


\begin{figure}[tbp]
\centering
\fbox{%
\begin{minipage}{0.95\textwidth}
\scriptsize
\setlength{\parskip}{2pt}
\setlength{\parindent}{0pt}

\textbf{Tool description.}

A tool for dynamic and reflective problem solving through thoughts. It supports flexible, evolving reasoning where each thought can build on, question, or revise previous insights.

\textbf{When to use:}
\begin{itemize}[itemsep=1pt, topsep=1pt, parsep=0pt]
  \item Complex problems that need stepwise reasoning
  \item Planning or analysis that may require revision
  \item Multi-step solutions or unclear initial scope
  \item Tasks needing context retention
  \item Situations requiring filtering of irrelevant info
\end{itemize}

\textbf{Key features:}
\begin{itemize}[itemsep=1pt, topsep=1pt, parsep=0pt]
  \item Adjustable \texttt{total\_thoughts}
  \item Ability to revise or question past thoughts
  \item Add thoughts even after reaching an apparent end
  \item Support for uncertainty, branching, and backtracking
  \item Hypothesis generation and verification
\end{itemize}

\textbf{Parameters:}
\begin{itemize}[itemsep=1pt, topsep=1pt, parsep=0pt]
  \item \texttt{thought} (string): The current thinking step
  \item \texttt{next\_thought\_needed} (boolean): Whether another thought step is needed
  \item \texttt{thought\_number} (integer): Current thought number
  \item  \texttt{total\_thoughts} (integer): Estimated total thoughts needed
  \item \texttt{is\_revision} (boolean, optional): Whether this revises previous thinking 
  \item  \texttt{revises\_thought} (integer, optional): Which thought is being reconsidered
  \item \texttt{branch\_from\_thought} (integer, optional): Branching point thought number 
  \item  \texttt{branch\_id} (string, optional): Branch identifier
  \item \texttt{needs\_more\_thoughts} (boolean, optional): If more thoughts are needed
\end{itemize}

\textbf{You should:}
\begin{enumerate}[itemsep=1pt, topsep=1pt, parsep=0pt]
  \item Start with an adjustable estimate of needed thoughts
  \item Revise previous reasoning when appropriate
  \item Add thoughts freely, even at the end
  \item Express uncertainty when relevant
  \item Mark revisions or branches
  \item Ignore irrelevant information
  \item Generate and verify hypotheses
  \item Iterate until satisfied
  \item Provide a single correct final answer
  \item Set \texttt{next\_thought\_needed} to false only when truly done
\end{enumerate}

\end{minipage}
}
\caption{Tool Description for the Sequential Thinking tool.}
\label{fig:sequential-thinking-tool-description}
\end{figure}

\begin{enumerate}
    \item \textbf{Purpose}: This component defines the tool's functional core and identity. It must clearly state \textit{what} the tool does, independent of specific task context. For example, in Figure~\ref{fig:sequential-thinking-tool-description}, the purpose is established in the opening statement: ``A tool for dynamic and reflective problem solving through thoughts.'' This primes the model's attention mechanism to the tool's fundamental capabilities. This component is mentioned in nine out of 15 sources along with the official documentation~\cite{towardsai-mcp, merge-dev-mcp-tools, conduit-github-mcp, speakeasy-mcp, medium-effective-mcp, reddit-good-mcp, xu2025llm, qin2023toolllm, anthropic-writing-tools-for-agents}. 
    
    \item \textbf{Guidelines}: This component addresses \textit{when} and \textit{how} the tool should be utilized. It provides decision-making criteria for activation and operational conduct for the FM. As shown in Figure~\ref{fig:sequential-thinking-tool-description}, this is distributed into two distinct logical blocks:
    \begin{itemize}
        \item \textit{Activation Criteria (When):} The ``When to use'' section explicitly lists appropriate task types (e.g., ``Tasks needing context retention'').
        \item \textit{Operational Instructions (How):} The ``You should'' section provides behavioral protocols (e.g., ``Start with an adjustable estimate'').
    \end{itemize}
    Consequently, if the guidelines are unclear, overly generic, or if explicit guidance is entirely missing, we classify this deficiency as a \textbf{\textit{Missing Usage Guidelines}} smell. This component is also mentioned in four sources~\cite{towardsai-mcp, writing-effective-tool, reddit-good-mcp, anthropic-writing-tools-for-agents} beyond the official documentation.
   
    \item \textbf{Limitations}: A tool description should describe known constraints, caveats, or corner cases where the tool may fail or be less effective. For example, a limitation of a calculator tool can be that it only supports up to two decimal point precision. We find this component is mentioned in only two sources apart from the official documents~\cite{writing-effective-tool, anthropic-writing-tools-for-agents}.
    
    \item \textbf{Parameter Explanation}: A tool description may include detailed explanations of all input parameters and their intended roles. In MCP, the input schema provides the mandatory structural specification for a tool’s parameters, including parameter names, data types, and required fields. Therefore, parameter explanation is evaluated with respect to this schema rather than only the free-form natural-language description. Figure~\ref{fig:sequential-thinking-tool-description} demonstrates this in the \textit{Parameters} section, where nine parameters are defined not just by data type (e.g., boolean, integer), but by intent (e.g., \texttt{is\_revision}: ``Whether this revises previous thinking''). This is the second-highest component in terms of mention, as we detect it in eight sources, in addition to official documents~\cite{towardsai-mcp, merge-dev-mcp-tools, conduit-github-mcp, writing-effective-tool, speedscale-mcp-tips, xu2025llm, qin2023toolllm, anthropic-writing-tools-for-agents}.
    
    \item \textbf{Length and Completeness}: A tool description should contain at least three to four sentences to ensure adequate detail. Complex tools warrant expanded descriptions; the Sequential Thinking tool (Figure~\ref{fig:sequential-thinking-tool-description}) necessitates a multi-section structure to fully capture its branching logic, validating the need for variable length based on complexity. Apart from the official documents, we find it in three sources~\cite{conduit-github-mcp, writing-effective-tool, reddit-mcp-best-practices}.  
    
    \item \textbf{Examples}: A tool description may include one or more illustrative examples demonstrating correct and effective usage. We observe example-related discussions in three sources beyond the official documents~\cite{merge-dev-mcp-tools,xu2025llm,qin2023toolllm}.
\end{enumerate}
\begin{figure}[t]
\centering
\begin{tcolorbox}[colback=gray!10, colframe=black!60, boxrule=0.8pt, arc=4pt, width=0.95\linewidth]
\textbf{Purpose:} How clearly and completely does the tool description explain what the tool does?

\begin{itemize}[leftmargin=1.2em, itemsep=2pt]
    \item \textbf{5/5:} Clearly explains function, behavior, and return data with precise language.
    \item \textbf{4/5:} Explains function and behavior with only minor ambiguity.
    \item \textbf{3/5:} Basic explanation present but lacks behavioral or output details.
    \item \textbf{2/5:} Vague or incomplete purpose statement.
    \item \textbf{1/5:} Purpose unclear or missing.
\end{itemize}

\end{tcolorbox}
\caption{Scoring instrumentation for the \textbf{Purpose} component. To ensure granular evaluation, this 5-point Likert scoring is applied independently to each of the six components.}
\label{fig:purpose-rubric}
\end{figure}
\subsubsection{Scoring \& smell derivation}\label{rubric-instrumentation}
While in~\secref{deriving-key-component} we identified the six core recommended components of tool descriptions, here we use these components to derive a scoring mechanism to measure the quality of tool descriptions. Prior studies indicate that FM-based evaluations become significantly more consistent and interpretable when guided by well-defined analytic rubrics rather than open-ended prompts~\cite{wang2025can, pathak2025rubric}. Following these insights, and adopting methodologies from prior work in automated FM-driven evaluation~\cite{wang2025can}, we implement a 5-point Likert scale~\cite{joshi2015likert} rubric for each component. This scale offers higher resolution than binary classification, allowing us to grade the quality of a component in a tool description from ``missing'' to ``ideal''. 


Since tool descriptions consist of unstructured natural language, evaluating the quality of these components is not a binary proposition. A component might be technically present in a tool description but semantically ambiguous or sub-optimal, making a simple ``Yes/No'' checklist insufficient. As a representative example, Figure~\ref{fig:purpose-rubric} details the specific performance descriptors for the \textit{Purpose} component. We designate score 3 as the \textbf{minimum threshold}: it represents a ``Minimum Viable'' description where the basic purpose is available. Scores 4 and 5 reward increasing precision, behavioral detail, and clarity. Conversely, scores 1 and 2 capture failure modes, reflecting descriptions that are vague, incomplete, or functionally sub-optimal. We apply this same rigorous scalar definition to all six identified components; the complete set of rubrics and associated evaluation prompts is provided in the Appendix~\ref{prompt-jury}.

\begin{figure}[t]
    \centering
    \includegraphics[width=0.8\linewidth]{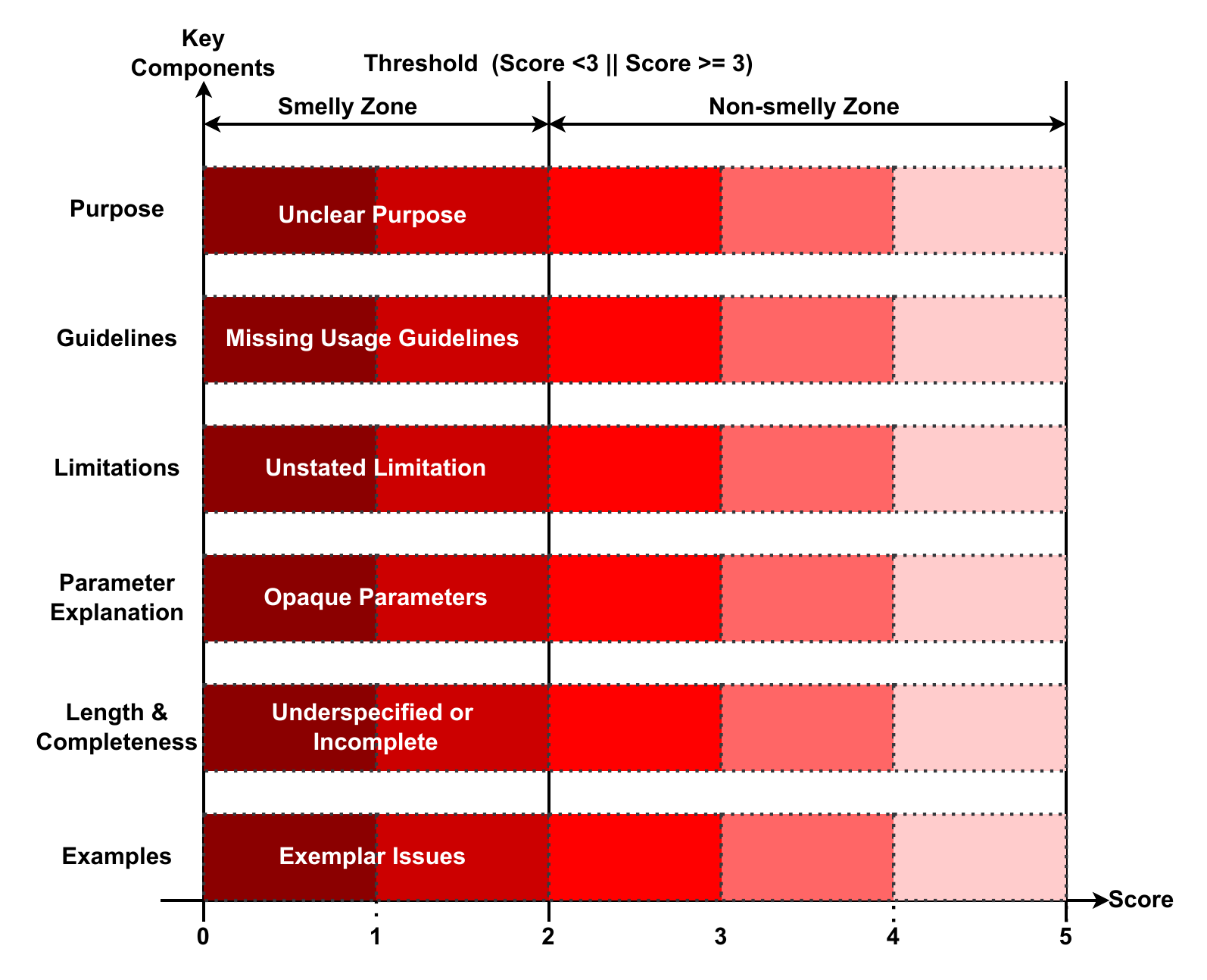}
    \caption{Mapping component scores to the tool description smells. The vertical threshold establishes Score 3 as the ``Minimum Viable'' standard, at which basic requirements for a given component are satisfied. The ``Smelly Zone'' (Scores $<3$) captures distinct smells, with the corresponding smell names shown within the zone, when a component is missing or inadequately specified, whereas the ``Non-smelly Zone'' (Scores $\ge 3$) indicates that the description contains the corresponding component, ranging from bare minimum presence (Score 3) to highly precise (Scores 4--5) specification.}
    \label{fig:smell-taxonomy-mapping}
    \label{fig:rubric-2-smell}
\end{figure}

We interpret scores below the minimum threshold (Score < 3) as indicators of qualitative deficiencies. Following prior work on design smells in specifications and prompts~\cite{moha2009decor, vogelsang2025impact}, we refer to recurrent, component-specific deficiencies in the tool descriptions as \textit{smells}. Figure~\ref{fig:rubric-2-smell} illustrates this mapping. For each component, scores in the smelly zone (Score < 3) directly indicate a corresponding smell: \textit{Unclear Purpose}, \textit{Missing Usage Guidance}, \textit{Unstated Limitation}, \textit{Opaque Parameters}, \textit{Underspecified or Incomplete}, and \textit{Exemplar Issues}. As each of these smells is deterministically derivable from the component and its score, we do not need any additional heuristic or classifier to identify the smells in the tool descriptions.

\subsection{Tool Description Collection}\label{tool-description-collection}
\subsubsection{MCP server curation}\label{dataset-curation} To ensure that our collection of MCP servers is both representative and useful, we conduct a lightweight literature search to identify prior studies that evaluate the execution of real-world complete workflows with MCPs. Following prior studies~\cite{hirsch2022systematic}, we perform a keyword-based search on Google Scholar using the queries \textit{``Evaluating MCP Servers''} and \textit{``MCP AND Benchmark''}. For each query, we sort the results by the relevance ranking provided by Google Scholar and manually inspect the first 100 results to identify studies that evaluate MCP servers.
For the first query, we identify 10 relevant articles, while for the second query, we identify 8 relevant articles. Since all 8 articles identified in the second query also appear within the 10 articles from the first query, our initial pool consists of 10 unique articles.

We then apply the following inclusion criteria to determine which studies provide MCP servers that are appropriate for our analysis:
\begin{itemize}
 \item The study uses real-world, MCP servers.
 \item The study evaluates MCP servers empirically or through benchmark analysis for general-purpose scenarios.
 \item The study publicly releases both the codebase and the MCP servers used.
 \end{itemize}

After applying these criteria, four studies meet all requirements~\cite{mo2025livemcpbench, luo2025mcp, fan2025mcptoolbench++, liu2507mcpeval}. From these studies, we curate 856 tools across 103 MCP servers reported in recent literature as of August 20, 2025. To support a comparative analysis of tool description smells between official and community-maintained MCP servers, we examined server documentation to identify their maintainers. Among the 103 MCP servers, we identify that 23 are maintained by official organizations, including Anthropic and major industry contributors such as GitHub, Airbnb, PayPal, and Microsoft, while the remaining 80 are maintained by the broader open-source community.
\subsubsection{Tool description extraction via MCP client}\label{tool-description-extraction}
We develop a lightweight MCP client to extract the tool descriptions from MCP servers. This MCP client dynamically communicates with MCP servers through the reflection protocol~\cite{hasan2025model} by sending a \texttt{tools/list} request. In response, each server returns a structured array containing all available tools along with their names, descriptions, and input schema. We implement this dynamic approach, avoiding static analysis, to ensure generalizability across the heterogeneous MCP ecosystem. Unlike static analysis, which relies on source code availability of MCP servers and language-specific parsers, this approach utilizes the native reflection of MCP to extract descriptions from any MCP server, regardless of its underlying implementation language (e.g., Python, Rust) or proprietary ownership status (e.g., open source or proprietary). The scanner then extracts the tool descriptions and input schema from the MCP client and prepares them for use in the subsequent evaluation step.

\subsection{Scanning Smells in the Tool Descriptions}\label{automated-evaluation-framework} 
After collecting tool descriptions from the MCP client, we evaluate their quality using an FM-based automated scanning framework grounded in the rubric developed in~\secref{rubric-development}. Because the consumers of tool descriptions are FMs themselves, assessing the quality of these descriptions requires an evaluator that reflects how FMs interpret and use them, making FM-based automated scanners a natural choice. However, FM evaluators introduce risks, including potential reproducibility issues and model-specific bias~\cite{bavaresco2025llms, wang2024large}. To address these challenges, we structure our scanning around three design principles: (i) a rubric-based scoring prompt, (ii) a controlled evaluation environment, and (iii) a multi-model LLM-as-jury configuration to improve robustness and generalization.

\subsubsection{Rubric-based evaluation prompt design}\label{rubric-guided-fm-evaluation} Prior studies have shown that FM-based techniques are effective at identifying structural issues and quality problems in software artifacts, including code smells~\cite{sadik2025benchmarking, wu2024ismell} and test smells~\cite{lucas2024evaluating}. Following these insights, we design a structured system prompt that operationalizes the scoring mechanism described in~\secref{rubric-instrumentation}. We use this prompt to evaluate each tool description collected from the MCP servers with an FM, which assigns an ordinal score between 1 and 5 for each component defined in~\secref{deriving-key-component}. As a result, each tool description is associated with six component-wise scores. We provide the full prompts and per-tool evaluations in the Appendix~\ref{prompt-jury}.

\subsubsection{Multi-model LLM-as-Jury evaluation} To mitigate FM bias and ensure generalizability across FM families, we adopt a multi-model LLM-as-jury configuration. Instead of relying on a single FM, we task three distinct FMs to independently execute the rubric-based scoring for every tool description in our dataset. Following prior study, we select models from three disparate families, e.g., \model{gpt-4.1-mini}, \texttt{claude-haiku-3.5}, and \texttt{qwen3-30b-a3b}, to ensure that our evaluation measures the quality of the description without any preferences specific to a single model architecture~\cite{verga2024replacing}.

To quantify the level of agreement among the models on the scores assigned, we compute the intraclass correlation coefficient (ICC), specifically ICC(2,1), which is widely used to measure absolute agreement among raters evaluating the same target~\cite{shrout1979intraclass}. ICC values between 0.5 and 0.75 indicate moderate reliability, values between 0.75 and 0.9 indicate good reliability~\cite{koo2016guideline}.
Table~\ref{tab:icc_rubric_components} summarizes the ICC values for each component. The results demonstrate substantial agreement across the majority of the components. We observed good reliability in 5 out of 6 components and moderate reliability in \textit{Examples} component. This variability is expected because the presence and quality of examples can depend on stylistic preferences or model-specific tendencies.

\begin{table}[t]
\centering
\small
\caption{Intraclass correlation coefficient ICC(2,1) computed for each rubric component based on scores independently assigned by three LLMs of the LLM-as-Jury across all 856 tool descriptions in our dataset. The results indicate good inter-rater agreement for most components and moderate agreement for the \textit{Examples} component, supporting the robustness of the evaluation.}
\label{tab:icc_rubric_components}
\begin{tabular}{
  >{\raggedright\arraybackslash}p{4cm}   
  >{\raggedleft\arraybackslash}p{1.5cm}  
}
\toprule
\textbf{Rubric component} & \textbf{ICC (2,1)} \\
\midrule
Purpose                   & 0.82 \\
Guidelines           & 0.85 \\
Limitations               & 0.84 \\
Parameter Explanation     & 0.90 \\
Length \& Completeness   & 0.76 \\
Examples                  & 0.62 \\
\bottomrule
\end{tabular}
\end{table}

\subsubsection{Manual Validation for Rubric-Based Scoring}
To further validate the rubric-based evaluation performed by the LLM-as-Jury, we conduct a manual evaluation on a representative sample of tool descriptions. To obtain a representative sample across our 856 studied MCP tools, we first apply Cochran’s formula~\cite{chaokromthong2021sample} with finite population correction to determine the required sample size. Similar to the prior study from Tosin et al.~\cite{fadahunsi2025generative} we use a 95\% confidence level ($Z = 1.96$), an estimated population proportion of $0.5$ ($p$), with a 10\% margin of error ($\varepsilon = 0.10$) in the following formula, which yields an adjusted sample size of $n = 87$. We then randomly sample 90 MCP tools (from our 856 studied MCP tools) without replacement to ensure representativeness across all tool types.

\[ \text{Sample size} = \frac{\dfrac{z^{2} \, p(1-p)}{\varepsilon^{2}}} {1 + \left( \dfrac{z^{2} \, p(1-p)}{\varepsilon^{2} N} \right)} \]

Next, the first and second authors independently evaluate the sampled tool descriptions using the same six-component rubric as the LLM-as-Jury, following the established manual validation approach described in prior studies~\cite{henriksson2025finerweb,ahmed2025can}. For each rubric component, we compare the human-assigned scores with the corresponding aggregated LLM-as-Jury scores using weighted Cohen's Kappa, which accounts for the ordinal nature of the 5-point Likert scale~\cite{cohen1968weighted}.~\tabref{tab:rubric_agreement} shows the weighted Cohen's Kappa scores ranging from 0.72 to 0.89 across the six rubric components, all above the 0.70 threshold for substantial agreement~\cite{williamson2012framework}. Therefore, we conclude that the LLM-as-Jury scores are well aligned with human rubric-based assessments.

\begin{table}[t]
\centering
\small
\caption{Agreement between human evaluators and LLM Jury based on Weighted Cohen Kappa for rubric-based tool descriptions. Values represent the agreement for each rubric component.}
\label{tab:rubric_agreement}
\begin{tabular}{l r}
\toprule
\textbf{Rubric} & \textbf{Weighted Cohen Kappa} \\
\midrule
Purpose & 0.85 \\
Guidelines & 0.72 \\
Limitations & 0.84 \\
Parameter Explanations & 0.89 \\
Length \& Completeness & 0.88 \\
Examples & 0.75 \\
\bottomrule
\end{tabular}
\end{table}
\subsubsection{Smell identification}
To transition from the raw multi-model scores to definitive smell assignments, we apply an aggregation and thresholding procedure. For each tool description and each component, we calculate the average of the three scores given by the FM evaluators and consider this as the consolidated score. Formally, for a component with scores \texttt{Score1, Score2, Score3}:
\[ \text{Smell Detected} \iff \frac{1}{N} \sum_{i=1}^{N} Score_{i} < 3 \]

Since the scoring mechanism is developed by keeping score 3 as the threshold for a well-formed component (as defined in~\secref{rubric-instrumentation}), any averaged score falling below this threshold indicates that the component is sub-optimal. When this condition is met, we assign the corresponding smell from the taxonomy in~\secref{rubric-instrumentation}. For example, for the \textit{Purpose} component of the tool \texttt{airbnb\_search}, the three evaluators assigned scores of 3, 2, and 3, yielding an average of approximately 2.7, which falls below the viability threshold and therefore triggers the \textbf{Unclear Purpose} smell. In contrast, for the tool \texttt{find\_nearby\_places}, all three evaluators assigned a score of 5 for the \textit{Purpose} component, yielding an average of 5.0 and indicating a clean, non-smelly description.


\subsection{Resolving Smells via Tool Component Augmentation}\label{optimizing-tool-description}
The objective of the tool description \rqtwonoun is to fix the smells identified in~\secref{automated-evaluation-framework} while retaining the original meaning and intent of the tool description. To achieve this, we design a semi-automated \rqtwoactor that combines rubric-based \rqtwonoun with an FM to generate refined, comprehensive, and factually consistent tool descriptions, being motivated by prior FM-based prompt and text optimization techniques~\cite{opsahl2024optimizing, khattab2023dspy}.

\subsubsection{Initial augmentation of components}\label{initial-augmentation}
In the first stage, the \rqtwoactor collects the original tool descriptions from every MCP server included in the MCP-Universe~\cite{luo2025mcp} and the input schema from MCP servers through an MCP client following the same procedure described in~\secref{tool-description-extraction}. It then applies the rubrics defined in~\secref{rubric-development} to guide the \rqtwonoun process. Using the \model{GPT-4.1-mini} model as FM, the \rqtwoactor automatically enhances each description by improving coverage across five rubric components: \textit{Purpose}, \textit{Guidelines}, \textit{Limitations}, \textit{Parameter Explanation}, and \textit{Length}. The outputs from this stage are referred to as \texttt{init\_augmented\_description}, as these outputs serve as input for the next refinement step. We intentionally exclude the \textit{Examples} component at this stage because the model cannot reliably generate factually grounded examples without execution traces, and doing so would risk introducing hallucinated or incorrect examples. This issue is addressed in the next stage, where examples are constructed from actual tool executions.

\subsubsection{Generation of the "Examples and Limitations" components}
Although the FM improves most rubric components effectively, it cannot reliably infer realistic Examples and the full set of Limitations without access to execution context. Generating these elements purely from prompts risks hallucination or factual inconsistency. To address this issue, we collect execution traces from realistic tool usage and use these traces to ground the generation of examples and limitations.

In our implementation, we use Claude Desktop~\cite{claude-desktop-mcp} with the Sonnet 4.5 model in two stages. First, we provide Claude Desktop with the tool description and input schema, without connecting to an actual MCP server, and prompt it to generate realistic natural-language tasks for the target tool. Second, we use a separate Claude Desktop interface configured as an MCP client with the corresponding MCP server connected. We then provide the generated task to this interface, allowing the FM to invoke the tool and complete the task through the MCP protocol. From the resulting interaction, we extract the tool-call arguments, tool responses, and execution outcomes as trace logs. Finally, we provide these trace logs back to the FM and instruct it to identify grounded examples and limitations based on the observed executions.

To ensure coverage of both successful and failing behaviors, we generate at least two tasks per tool:
(i) at least one task that should result in a successful tool execution and produce a valid response, and
(ii) at least one task that should result in an empty response or an error. Additionally, we generate more tasks to cover edge cases, ranging from 1 to 3 depending on the complexity of the tool.

As an illustrative example, for the \texttt{get\_historical\_stock\_prices} tool in the \texttt{yfinance} MCP server, we construct the following tasks:

\begin{enumerate}
\item Retrieve Salesforce (CRM) prices from 1 Nov 2020 to 2 Dec 2020.

\item Retrieve Salesforce (CRM) prices for the same day, 2 Dec 2020.

\item Retrieve price changes from 10 Jan 2023 to 25 Jan 2025 and make multiple tool calls if needed.
\end{enumerate}

The first task produces a successful response and serves as a positive example. The second task produces an empty response because the tool requires a non-zero date range and therefore serves as a negative example. The third task encourages the FM to issue multiple calls due to the large date range, which helps capture multi-step behavior useful for identifying limitations related to response size and rate limits. The actual analysis steps in this pipeline, i.e., task generation, tool invocation, and identification of examples and limitations, are fully automated and performed by FMs and MCP-compatible clients. The only manual portion in this setup was boilerplate orchestration for transferring tasks between Claude Desktop interfaces and typing the resulting trace logs into a structured JSON file. In the following section, we describe a fully automated orchestration workflow that removes the manual transfer of tasks and logs.

\subsubsection{Automated Task Generation for Scalable Augmentation}\label{automated-task-generation}
After completing the evaluation, we automate the remaining manual steps using a fully automated workflow, which we include in the replication package for others to use. This workflow directly connects to MCP servers, retrieves each tool's description and input schema through the MCP client, and uses these artifacts to generate realistic natural-language tasks for the target tools. It is designed to produce tasks at multiple difficulty levels, prioritizing harder cases such as edge cases, boundary values, near-limit inputs, and unusual parameter combinations before easier standard-use tasks. We provide \model{GPT-4.1-mini} with the tool description and input schema, then prompt it to generate a realistic task for the target MCP tool programmatically. The generated task is then executed by a ReAct agent connected to the corresponding MCP server. After execution, the agent passes the original task and the execution trace to a Jury LLM, implemented with \model{GPT-4.1}, which extracts grounded Examples and Limitations from the observed behavior. This automated workflow eliminates the need for manual task transfer and log extraction while preserving the same grounding principle as our manually orchestrated setup. Practitioners may either manually create tasks for tools they know well or adopt this automated task generation framework to scale the augmentation process across larger MCP servers. We include the prompts used for task generation, execution, and jury-based extraction in the Appendix~\secref{augmentor-prompt}.

\subsubsection{Final consolidation of all components}\label{final-consolidation}
In the final stage, we feed the \texttt{init\_augmented\_description}, along with the collected JSON logs, into the \rqtwoactor FM. We instruct the FM to output a structured JSON object comprising five explicit fields: \textit{Purpose}, \textit{Guidelines}, \textit{Limitations}, \textit{Parameter Explanation}, and \textit{Examples}, mapping to distinct components of the tool descriptions as identified in~\secref{deriving-key-component}. 

We do not include a separate field for \textit{Length and Completeness}. This component functions as a meta-quality dimension of the overall tool description and is automatically fulfilled when the other five components are properly populated. Therefore, while it remains an essential dimension during scoring and smell detection, it does not require an independent field in the \rqtwoactionpast representation. Consequently, the resulting five-component \rqtwoactionpast tool descriptions are used directly in subsequent evaluations and ablation studies.

\subsection{Evaluating the Augmented Tool Descriptions}

\subsubsection{Benchmark adoption} We use the MCP-Universe benchmark~\cite{luo2025mcp} for evaluating the performance of agents with \rqtwoactionpast tool descriptions. We chose this benchmark because of its comprehensive design that integrates real-world scenarios and temporal dynamics. It includes 231 complex real-world tasks across six domains, while providing a total of 202 tools. Unlike several other benchmarks that rely solely on LLM-as-Jury evaluations~\cite{mo2025livemcpbench}, MCP-Universe combines this with a robust execution-based evaluation mechanism. In this benchmark, task outcomes are assessed by automated evaluators that directly execute tools and verify results against ground-truth criteria. These evaluators are programmatic validation scripts defined to verify format compliance, static content matching, and dynamic validation for temporally sensitive tasks. This combination enables a more accurate and reliable assessment of agent performance and ensures fair comparison across models. Each task is evaluated by at least one evaluator, with an average of 3.3 evaluators per task. 

To assess whether the 202 tools from the MCP-Universe benchmark have significantly different quality characteristics (e.g., higher or lower quality than the remaining 654 tools), we compare the tool description component scores of the MCP-Universe tools against those of the remaining 654 studied MCP tools. For each rubric component, we construct two independent score distributions: one containing the scores of the MCP-Universe tools, and another containing the scores of the remaining 654 studied MCP tools. Since the scores are ordinal and not normally distributed, we use a Mann-Whitney U test~\cite{mann1947test} for each component. As this results in six component-wise tests, one for each rubric component, we apply Bonferroni correction~\cite{armstrong2014use} across the six resulting p-values. We also report Cliff’s delta~\cite{macbeth2011cliff} to quantify the practical magnitude of the difference, using the standard thresholds of negligible ($|\delta| < 0.147$), small ($|\delta| < 0.33$), medium ($|\delta| < 0.474$), and large ($\delta > 0.474$) following the prior study by João et al.~\cite{bernardo2023impact}.

As shown in~\tabref{tab:rubric_mcp_vs_rest}, four of the six rubric components show no statistically significant difference after Bonferroni correction: Usage Guidelines, Limitations, Parameter Explanation, and Examples. Purpose and  Length \& Completeness show statistically significant differences, but their Cliff’s delta values are small or negligible ($\delta=-0.15$ and $\delta=-0.14$, respectively). Overall, these results indicate that the MCP-Universe subset is broadly comparable to the remaining corpus in tool-description quality.

\begin{table}[t]
\centering
\small
\caption{Comparison of tool description component scores between MCP-Universe benchmark tools and the remaining corpus using the Mann--Whitney U test. \textit{MU-Med.} denotes the median score of MCP-Universe tools, and \textit{R-Med.} denotes the median score of the remaining corpus. U-statistics are reported in thousands. Bonferroni correction is applied across the six rubric components, and effect sizes are measured using Cliff's delta for the ones where statistical difference is observed.}
\label{tab:rubric_mcp_vs_rest}
\begin{tabular}{l r r r r r r l}
\toprule
\textbf{Rubric} & \textbf{MU-Med.} & \textbf{R-Med.} & \textbf{U (k)} & \textbf{p} & \textbf{Adj. p} & \textbf{Cliff's $\delta$} & \textbf{Effect} \\
\midrule
Purpose                & 2.00 & 2.33 & 54.92 & <0.05 & <0.05 & -0.15 & small   \\
Usage Guidelines       & 1.00 & 1.00 & 58.01 & <0.05 & 0.08  & N/A & N/A \\
Limitations            & 1.00 & 1.00 & 58.03 & <0.05 & 0.08  & N/A & N/A \\
Parameter Explanation  & 1.00 & 1.00 & 62.52 & 0.32  & 1.00  & N/A & N/A \\
Examples               & 1.00 & 1.00 & 57.98 & <0.05 & 0.08  & N/A & N/A \\
Length \& Completeness                 & 1.33 & 1.67 & 55.93 & <0.05 & <0.05 & -0.14 & negligible\\
\bottomrule
\end{tabular}
\end{table}

\subsubsection{Tool Description Router}\label{mcp-client-calibration}
The original MCP client provided in MCP-Universe does not support dynamic modification of tool descriptions at runtime, which is required for our evaluation. To address this limitation, we extend the client with a configurable switching module called the Tool Description Router, which allows for the dynamic selection of tool descriptions. This module can load either the original descriptions or the \rqtwoactionpast descriptions stored in the PostgreSQL database described in~\secref{final-consolidation}, depending on a configuration provided as an argument.

Additionally, to support the ablation study of individual rubric components, the router also supports retrieving certain components in a specified order from the \rqtwoactionpast tool description and assembles a valid description for model consumption. For example, suppose we need to run the benchmark with only the \textit{Purpose} and \textit{Guideline} components of the \rqtwoactionpast tool description. In that case, those two component names can be passed as comma-separated arguments to the tool description router, which will fetch only these two components for all tools from the database, concatenate them, and present them to the FM.

\subsubsection{Full Rubric Evaluation}\label{full-rubric-evaluation}
Using the tool description router, we evaluate the \rqtwoactionpast tool descriptions by running the MCP-Universe benchmark with all rubric components retrieved from the database. 
Each task in MCP-Universe is associated with one or more evaluators that determine task completion correctness, as described in~\secref{sec:background}. To measure the impact of \rqtwoactionpast tool descriptions, we compare model performance against the baseline results reported in the original MCP-Universe study. We use three primary evaluation metrics:

\begin{itemize}
\item \textbf{Success Rate (SR)}, which measures the percentage of tasks that pass all evaluators;
\item \textbf{Average Evaluator score (AE)}, which represents the average fraction of evaluators that pass across all tasks within a domain or model; and
\item \textbf{Average \# of Steps (AS)}, which captures the average number of steps required for an FM to complete each task. These steps reflect how the agent leverages FM and tools in real-time while solving the task, e.g., the number of calls to FMs made by the agent.
\end{itemize}

To illustrate these metrics, consider two example tasks: the first task takes four steps to complete and is evaluated by three evaluators and passes all of them, while the second task takes six steps and is evaluated by four evaluators and passes only two. Hence, the SR is 50\% since one of the two tasks (task-01) passed all its evaluators. The proportion of evaluators passed for task-01 is 1.0 and for task-02 is 0.5, giving an AE of 0.75. This indicates that although only 50\% of the tasks fully passed, some evaluators in the failed tasks still passed as successful. Finally, AS is 5, calculated as the mean of four and six steps. Following prior similar studies~\cite{kapoor2024ai}, we analyze the accuracy-cost tradeoff of the \rqtwoactionpast tool descriptions using a Pareto curve that uses AS as a proxy for the cost metrics across all six domains for each model.

As there are 231 tasks in the MCP-Universe benchmark, running all the tasks against each model is costly in terms of token cost and time. Running the entire benchmark for one round with one FM can cost us around 200 to 300 million tokens, which translates to $75$ to $600$ USD, depending on the model provider. Hence, to balance the cost and generalization of our evaluation, we conduct experiments on one proprietary model (\model{GPT-4.1}) and two open-weight models (\model{Qwen3-Coder-480B-A35B} and \model{GLM-4.5 355B A32B}), following the configurations established in the original MCP-Universe benchmark~\cite{luo2025mcp}.

A limitation is that MCP-Universe does not report success information, the number of passed evaluators, or the number of steps at a per-task level, which prevents paired statistical tests. Re-executing these baselines to obtain per-task data would be prohibitively expensive. To overcome this, we adopt a hybrid strategy: (i) for the three original models (from MCP-Universe study~\cite{luo2025mcp}), we compare our \rqtwoactionpast results directly against the reported aggregated (average/percentage) baselines from MCP-Universe; and (ii) to enable more rigorous statistical comparison, we introduce a smaller-sized open-weight model, \model{Qwen3-Next-80B-A3B-Instruct}. We select this model for its cost-efficiency (0.10 USD per 1M tokens vs. 0.456 USD for \model{Kimi-K2}~\cite{kimi-k2-hf}) and strong performance on external benchmarks such as LiveCodeBenchv6 (56.56 versus 53.7)~\cite{qwen3-80b-next-hf, kimi-k2-hf}. For this model, we execute both the baseline (original descriptions) and treatment (\rqtwoactionpast descriptions) runs, generating the paired per-task data necessary for significance testing. 
 
Finally, we had to apply specific adaptations to accommodate model and domain constraints. As the smaller open-source model, i.e., \model{Qwen3-Next-80B-A3B-Instruct}, has a shorter context window, we use \textit{Purpose}, \textit{Guidelines}, and \textit{Limitation} components from the \rqtwoactionpast tool description for this model to avoid overloading the context window of this model. We exclude the \textit{Parameter Explanation} component as a context-length tradeoff. This exclusion remains feasible because the MCP protocol automatically provides the input schema at runtime, including parameter names and types, which preserves the minimum structural information required for tool invocation. Similarly, out of the six domains of the MCP universe, we avoid generating examples for the \texttt{Browser Automation} domain as the examples for the tools of this domain are too large to fit in the context window. Finally, the original MCP-Universe study utilizes the SERP API-based Google search MCP server, which provides 250 free queries per month per API key. As the Web Searching domain has 55 tasks, each of which requires multiple searches, it is not suitable for an overall study. Hence, we adopt the Google search MCP server~\footnote{\url{https://github.com/mixelpixx/Google-Search-MCP-Server}}, which uses the Google search API key to run the experiments uninterrupted. 

\subsubsection{Ablation Study} Since the \rqtwoactionpast tool description stored in the database contains the components \textit{Purpose}, \textit{Guidelines}, \textit{Limitations}, \textit{Parameter Explanation}, and \textit{Examples} (as explained in~\secref{final-consolidation}), we conduct an ablation study to determine which components contribute most significantly to performance. For example, Anthropic~\cite{anthropic-tool-description} considers the \textit{Examples} component less critical, whereas other studies suggest that it may be beneficial~\cite{xu2025llm}. Similarly, the \textit{Parameter Explanation} component may contain redundant information that overlaps with the input schema. Moreover, including all components may increase the input context length of the FM and affect efficiency.

To investigate these effects, we use the \texttt{--components} command of the tool description router to selectively enable different combinations of rubric elements. We design two experimental configurations:
(i) excluding the \textit{Examples} component while retaining \textit{Purpose}, \textit{Guidelines}, \textit{Limitation}, and \textit{Parameter Explanation} to measure the effect of examples; and
(ii) evaluating pairwise combinations of two components, where one component is always \textit{Purpose} (for instance, \texttt{Purpose + Limitation}, \texttt{Purpose + Parameter Explanation}, and \texttt{Purpose + Examples}). We include \textit{Purpose} in all combinations because it defines what the tool does, and without it, the FM cannot correctly infer the tool’s intent or functionality.

\section{Results}\label{sec:results}

\subsection{RQ-1: To what extent do MCP tools' descriptions contain smells?}\label{results-rq1}
\motivation
Writing MCP tool descriptions requires combining principles from software requirements specification and prompt engineering, suggesting that these descriptions may inherit suboptimal design patterns or smells from both domains. Historically, smells in software engineering have been known to be related to change-proneness and bugs~\cite{khomh2009exploratory,hassan2022code}, motivating an investigation of tool description smells in MCP. Although 66\% of MCP servers already exhibit code-level smells~\cite{hasan2025model}, it remains unclear whether similar issues manifest in the natural-language descriptions that guide agent behavior.

\approach
We develop a rubric for evaluating MCP tool descriptions which consists of the components of tool description derived from Anthropic’s design guidelines, practitioners' recommendations, and prior research, as well as a structured scoring scale as mentioned in~\secref{rubric-development}. As FMs are the primary consumers of the tool description, to evaluate whether FMs can understand and interpret these tool descriptions, we score the 856 tools collected from MCP servers through an FM-based scanner. This scanner consists of an LLM-as-Jury with 3 FMs using the rubric, as detailed in~\secref{automated-evaluation-framework}. Then we identify the cases with low scores and map those to recurring smells. Given prior work showing differences between official and community-maintained MCP servers~\cite{hasan2025model}, we conduct pairwise nonparametric comparisons using the Mann–Whitney U test and refine the results with Bonferroni-adjusted $p$-values~\cite{ruxton2008time} to uncover any such differences.

\begin{figure}[t]
    \centering
    \includegraphics[width=0.9\textwidth]{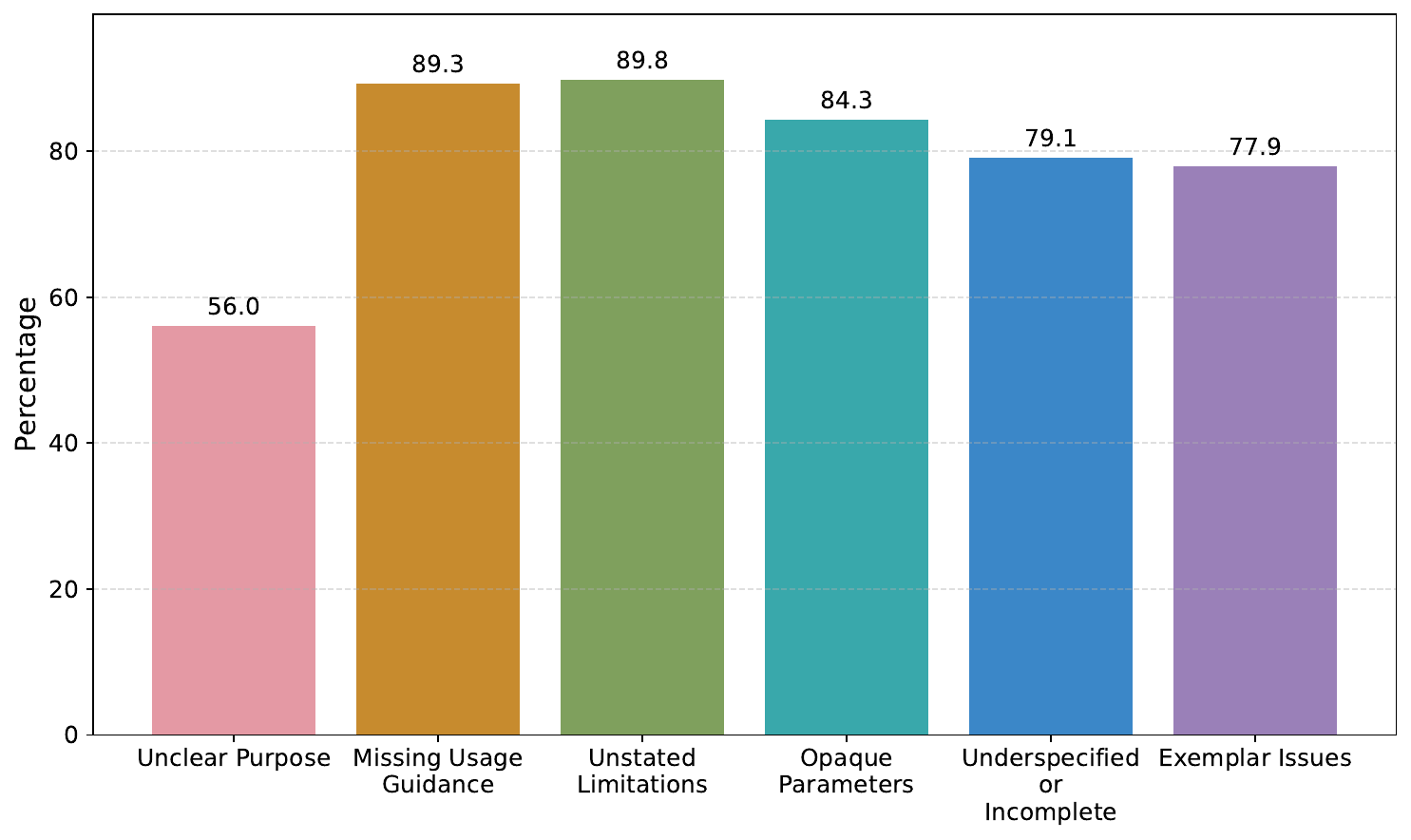}
    \caption{Prevalence of smell types in the tool descriptions of MCP servers.}
    \label{fig:rq1:smell_prevalence}
\end{figure}

\findings
\textbf{All six smell types affect the majority of MCP tool descriptions, with the most severe issues appearing in nearly 90\% of tools}. As shown in Figure~\ref{fig:rq1:smell_prevalence}, the most widespread smell categories are Unstated Limitations (89.8\%), Missing Usage Guidelines (89.3\%), and Opaque Parameters (84.3\%). These patterns suggest that tool descriptions frequently lack critical boundary conditions, fail to indicate when or how tools should be invoked, and provide little insight into the meaning or behavioral implications of input parameters. The next tier includes Underspecified or Incomplete descriptions (79.1\%) and Exemplar Issues (77.9\%), which arise when descriptions are overly brief relative to tool complexity or rely on sparse or uninformative examples instead of clear explanatory text. Even for the best-performing component (i.e., Purpose), we observe the Unclear Purpose smell in 56\% of tools, indicating that more than half of tool descriptions do not clearly articulate their intended functionality. Taken together, these suboptimal patterns can hinder the ability of FMs to properly solve real-world problems. Prior work shows that under-specified prompts can lead to up to 2$\times$ performance regressions across model versions~\cite{yang2025prompts}, suggesting that such pervasive tool description smells may similarly increase brittleness and reduce reliability in MCP-enabled agent behaviors.


We illustrate a high-quality description of a tool in Figure~\ref{fig:sequential-thinking-tool-description} (presented in~\secref{deriving-key-component}) from the Sequential Thinking MCP server. The description clearly conveys the tool’s purpose, guidelines, limitations, and parameters. In contrast, Figure~\ref{fig:mcp-poor-tooldesc-examples} shows three examples of low-quality tool descriptions that scored below the median quality score. These minimal descriptions provide little contextual information, leaving foundation models with insufficient cues to infer when or how to use the tools effectively.

\textbf{Only 2.9\% of MCP tool descriptions are fully smell-free}. As shown in Table~\ref{tab:smell_free_combinations}, the number of smell-free instances drops sharply when analyzing tool descriptions with larger combinations of components considered together. While certain individual components have relatively high smell-free rates, e.g., 44.0\% for \textit{Purpose} alone and 10.4\% with \textit{Guidelines}, the proportion declines to 7.5\% when analyzing tools that combine \textit{Purpose}, \textit{Guidelines}, and \textit{Limitations} components in their description, and drops to a mere 2.9\% when all five components in the description are required to be smell-free. This pattern indicates that many tool descriptions are highly incomplete in terms of components.

\begin{table}[t]
\centering
\caption{Smell-free tool description counts and percentages across rubric combinations. Notation: P = Purpose; G = Guidelines; L = Limitation; PEx = Parameter Explanation; E = Examples.}
\label{tab:smell_free_combinations}
\begin{tabular}{
lrr
}
\toprule
\textbf{Rubric combination} & \textbf{\# Smell-free} & \textbf{\% Smell-free} \\
\midrule
P                           & 376 & 44.0 \\
P + G                       & 89  & 10.4 \\
P + G + L                   & 64  & 7.5 \\
P + G + L + PEx             & 26  & 3.0 \\
P + G + L + PEx + E         & 25  & 2.9 \\
\bottomrule
\end{tabular}
\end{table}

\begin{figure}[t]
    \centering
    \includegraphics[width=0.99\textwidth]{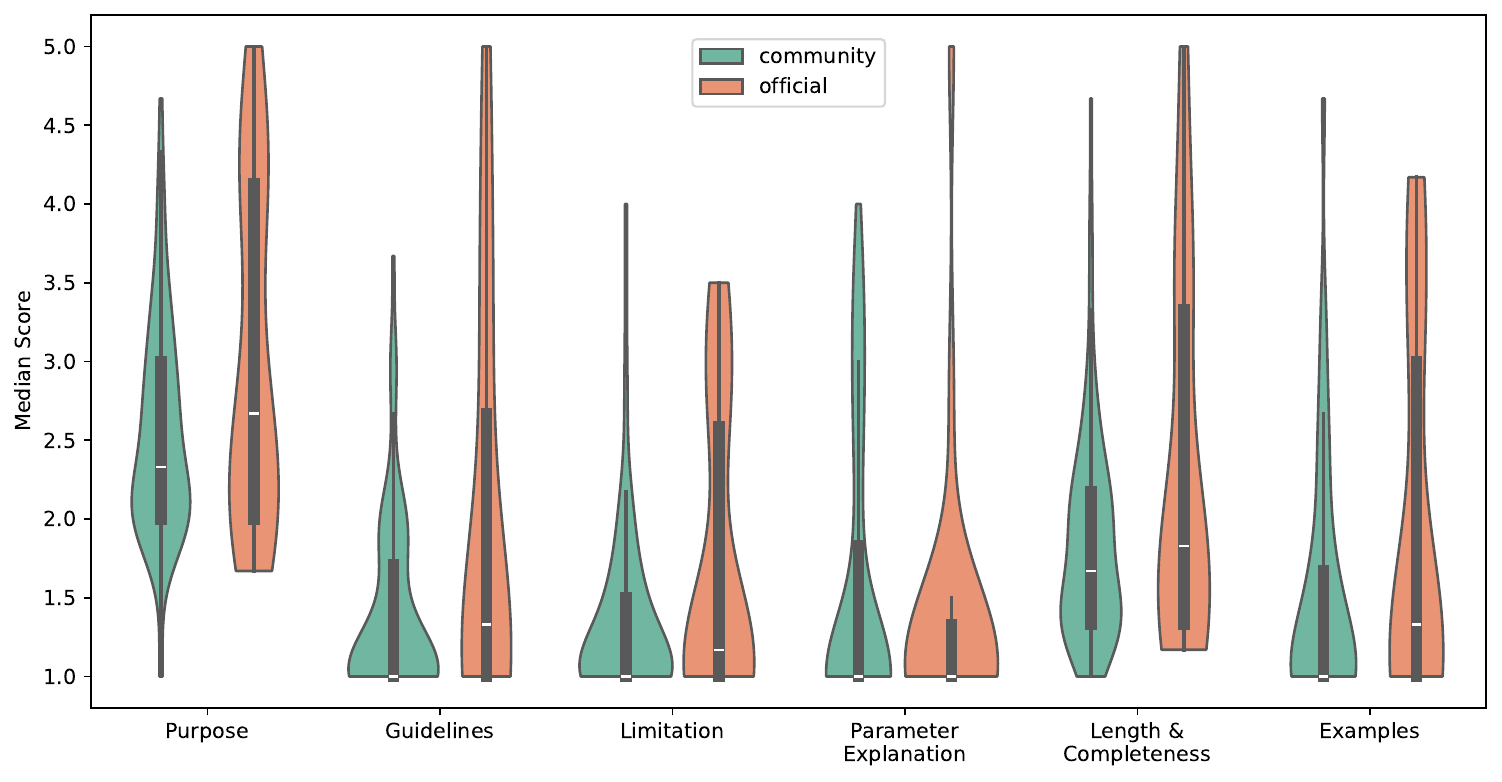}
    \caption{Distribution of the median scores across the six components of the rubric among the official and community MCP servers.}
    \label{fig:rq1:violin_plot_quality_score}
\end{figure}

\begin{table}[t]
\centering
\small
\caption{Statistical test results comparing median component scores between community and official MCP servers using the Mann--Whitney U test, with Bonferroni correction applied to p-values.}
\label{tab:mann_whitney_official_vs_community}
\begin{tabular}{
p{3.0cm}  
r         
r         
r         
p{2.6cm}  
}
\toprule
\textbf{Score}  &
\textbf{Statistic} & \textbf{p-value} & \textbf{Adj. p-value} \\
\midrule
Purpose  &  873.50 & 0.18 & 1.00 \\
Guidelines  &  759.50 & 0.17 & 1.00\\
Limitations &  824.00 & 0.42 & 1.00\\
Parameter Explanation  & 973.00 & 0.61 & 1.00\\
Length \& Completeness  & 829.00 & 0.46 & 1.00\\
Examples & 783.00 & 0.24 & 1.00\\
\bottomrule
\end{tabular}
\end{table}

\textbf{Official and community MCP servers show low-quality tool descriptions across all components of our rubric}. We visualize the distributions of median scores for each rubric item across official and community-maintained MCP servers in Figure~\ref{fig:rq1:violin_plot_quality_score}. To compare the two groups, we apply the Mann-Whitney U test with Bonferroni correction for the six components. As summarized in Table~\ref{tab:mann_whitney_official_vs_community}, none of the components show statistically significant differences between official and community servers; all raw p-values exceed 0.17 and all corrected p-values equal 1.0.

\begin{figure}[tbp]
\centering
\fbox{%
\begin{minipage}{0.9\textwidth}
\scriptsize
\setlength{\parskip}{2pt}
\setlength{\parindent}{0pt}

\medskip

\textbf{Tool name:} \texttt{create\_invoice} (official)\\
\textbf{Description:} Creates PayPal Invoice Link.

\medskip

\textbf{Tool name:} \texttt{read\_mail}(community)\\
\textbf{Description:} Retrieves the content of a specific email.

\medskip

\textbf{Tool name:} \texttt{maps\_place\_details(community)}\\
\textbf{Description:} Get detailed information about a specific place.

\end{minipage}
}
\caption{Example MCP tool descriptions with low quality score.}
\label{fig:mcp-poor-tooldesc-examples}
\end{figure}

\begin{Summary}{Summary of RQ-1}{

\begin{enumerate}

    \item MCP tool descriptions exhibit a pervasive lack of quality, with \textbf{97.1\%} containing at least one description smell.
    \item Even the least prevalent smell, \textit{Unclear Purpose}, appears in \textbf{56\%} of tools, indicating that over half fail to state their core functionality clearly.
    \item Moreover, official and community tools reveal no statistically significant differences, suggesting that poor quality is a systemic issue within the ecosystem.
\end{enumerate}
}
\end{Summary}

\subsection{RQ-2: How does resolving tool description smells by \rqtwoaction all tool description components impact the performance of FM-based agents?}\label{results-rq2}

\motivation
The pervasive presence of smells in tool descriptions (as found in RQ-1) naturally motivates attempts to resolve them; however, theoretical signals regarding their impact are conflicting. While software engineering research warns that smell removal can unintentionally alter system behavior~\cite{verdecchia2018empirical}, prompt engineering studies suggest that richer descriptions can improve outcomes~\cite{khattab2023dspy}. We therefore investigate whether systematically resolving tool description smells by \rqtwoaction all components yields a net positive impact on agent performance on standard benchmarks.

\approach
Following the steps mentioned in \secref{optimizing-tool-description}, we \rqtwoverb all components of the tool description, resulting in a fully \rqtwoactionpast description that includes \textit{purpose}, \textit{guidelines}, \textit{limitations}, \textit{parameter explanation}, and \textit{examples}. As discussed in~\secref{final-consolidation}, \textit{length and completeness} are implicitly addressed through these components rather than \rqtwoactionpast independently. After \rqtwoaction the tool descriptions, we again measure the quality score of the fully \rqtwoactionpast tool descriptions using the same multi-model LLM-as-Jury as in \secref{automated-evaluation-framework} to validate whether the fully \rqtwoactionpast tool descriptions achieve a higher quality score. We further apply the Wilcoxon signed-rank test~\cite{woolson2007wilcoxon} to determine whether the observed improvements in quality scores are statistically significant, as this test detects systematic shifts in the central tendency of paired observations in before and after \rqtwonoun.

In addition, we test whether the observed changes in SR, AE, and AS are statistically significant. As the original study did not provide these metrics for every (task, model) combination, we measure these metrics using the model introduced in our work, i.e., \model{Qwen3-Next-80B-A3B-Instruct}. Since SR is a binary outcome (0/1), we assess the significance of SR changes using McNemar’s test in its chi-squared formulation~\cite{pembury2020effective}. In contrast, as AE and AS are continuous-valued metrics, we apply a Wilcoxon signed-rank test to determine their statistical significance.

\begin{table}[t]
\centering
\small
\caption{Wilcoxon signed-rank test results comparing tool description component scores before \rqtwonoun (BA) and after \rqtwonoun (AA), showing statistically significant median score increases across all components.}
\label{tab:wilcoxon_bo_ao_full}
\begin{tabular}{lrrrrr}
\toprule
\textbf{Component} &
\textbf{Statistic} &
\textbf{p-value} &
\textbf{Med. Score (BA)} &
\textbf{Med. Score (AA)} &
\textbf{Med. diff.} \\
\midrule
Purpose                   & 0.0 & $<0.001$ & 2.0 & 5.0 & 2.7 \\
Guidelines           & 0.0 & $<0.001$ & 1.0 & 5.0 & 4.0 \\
Limitations                & 0.0 & $<0.001$ & 1.0 & 5.0 & 3.7 \\
Parameter explanation     & 0.0 & $<0.001$ & 1.0 & 4.7 & 3.3 \\
Examples                  & 0.0 & $<0.001$ & 1.0 & 5.0 & 3.7 \\
Length and completeness   & 0.0 & $<0.001$ & 1.3 & 5.0 & 3.7 \\
\bottomrule
\end{tabular}
\end{table}

\findings
\textbf{FM-based \rqtwonoun resolves the tool description smells in the MCP-Universe benchmark by improving the median scores by 2.7--4.0 points in our scoring rubric}. We perform the Wilcoxon signed-rank test on each tool description to compare scores before (BA) and after \rqtwonoun (AA) and report the findings in \tabref{tab:wilcoxon_bo_ao_full}. These results confirm statistically significant improvements across all six components ($p < .001$). We observe that for every component, the test statistic equals zero, establishing that all paired comparisons favor the \rqtwoactionpast descriptions, with median scores rising by 2.7 to 4.0 points on the five-point Likert scale. We find that median BA scores cluster between 1.0 and 2.0, reflecting widespread pre-\rqtwonoun under-specification, whereas median AA scores converge near the ceiling value of 5.0 for all components.

\begin{table}[t]
\centering
\small
\caption{Success rate (SR) comparison of models using original vs.\ \rqtwoactionpast tool descriptions across six domains. Original SR values for \model{GPT-4.1}, \model{Qwen3-Coder-480B-A35B}, and \model{GLM-4.5} are taken from the MCP-Universe baseline study~\cite{luo2025mcp}, whereas Original SR for \model{Qwen3-Next-80B-A3B-Instruct} is obtained by running the agent with the original tool descriptions. The $\Delta$SR column reports the absolute change in success rate in percentage points (SR$_{\text{after \rqtwonoun}}$ -- SR$_{\text{original}}$). The statistically significant improvements in description quality (observed in~\tabref{tab:wilcoxon_bo_ao_full}) also translate into higher task success rates in more than half of the domain-model combination rows, while also resulting in performance regressions in a smaller subset (16.67\%) of cases.}
\label{tab:optimized_sr_comparison}
\begin{tabularx}{\textwidth}{
  >{\raggedright\arraybackslash}p{2.6cm}
  >{\raggedright\arraybackslash}p{2.8cm}
  >{\raggedleft\arraybackslash}p{2.0cm}
  >{\raggedleft\arraybackslash}p{2.0cm}
  >{\raggedleft\arraybackslash}p{1.8cm}
}
\toprule
\textbf{Model name} & \textbf{Domain} & \textbf{Original SR} & \textbf{SR after \rqtwonoun} & \textbf{$\Delta$SR (pp)} \\
\midrule
\multirow{6}{=}{\model{GPT-4.1}}
& Finance             & 40.00\% & 57.50\% & \cellcolor{green!25}17.50\% \\
& Repo Management     & 6.06\%  & 21.20\% & \cellcolor{green!25}15.14\% \\
& 3D Design           & 26.32\% & 31.60\% & \cellcolor{green!25}5.28\%  \\
& Location Navigation & 8.89\%  & 31.00\% & \cellcolor{green!25}22.11\% \\
& Browser Automation  & 23.08\% & 25.65\% & \cellcolor{green!25}2.57\% \\
& Web Searching       & 10.91\% & 10.91\% & 0.00\% \\
\midrule
\multirow{6}{=}{\model{Qwen3-Coder-480B-A35B}}
& Finance             & 40.00\% & 72.50\% & \cellcolor{green!25} 32.50\% \\
& Repo Management     & 3.03\%  & 18.20\% & \cellcolor{green!25}15.17\% \\
& 3D Design           & 26.32\% & 21.10\% & \cellcolor{red!20}{-5.22}\% \\
& Location Navigation & 8.89\%  & 15.60\% & \cellcolor{green!25}6.71\% \\
& Browser Automation  & 25.64\% & 23.07\% & \cellcolor{red!20}{-2.57}\% \\
& Web Searching       & 10.91\% & 9.10\%  & \cellcolor{red!20}{-1.81}\% \\
\midrule
\multirow{6}{=}{\model{GLM-4.5}}
& Finance             & 50.00\% & 67.50\% & \cellcolor{green!25}17.50\% \\
& Repo Management     & 9.09\%  & \textit{Could not run} & \textit{N/A} \\
& 3D Design           & 26.32\% & 26.32\% & 0.00\% \\
& Location Navigation & 17.78\% & 17.78\% & 0.00\% \\
& Browser Automation  & 15.38\% & 15.38\% & 0.00\% \\
& Web Searching       & 27.27\% & 18.18\% & \cellcolor{red!20}{-9.09}\% \\
\midrule
\multirow{6}{=}{\model{Qwen3-Next-80B-A3B-Instruct}}
& Finance             & 50.00\% & 65.00\% & \cellcolor{green!25}15.00\%  \\
& Repo Management     & 18.18\% & 18.18\% & 0.00\% \\
& 3D Design           & 0.00\%  & 10.53\% & \cellcolor{green!25}10.53\% \\
& Location Navigation & 11.11\% & 13.33\% & \cellcolor{green!25}2.22\% \\
& Browser Automation  & 12.82\% & 12.82\% & 0.00\% \\
& Web Searching       & 0.00\%  & 7.27\%  & \cellcolor{green!25}7.27\%  \\
\midrule
\multicolumn{4}{r}{\textbf{Cross-Model Median $\Delta$SR}} & \textbf{5.85\%} \\
\multicolumn{4}{r}{Improved cases \textbf{($\Delta$SR $>$ 0)}} & \textbf{54.17\%} \\
\multicolumn{4}{r}{Regressed cases \textbf{($\Delta$SR $<$ 0)}} & \textbf{16.67\%} \\
\multicolumn{4}{r}{Unchanged or failed runs} &  \textbf{29.16\%} \\
\bottomrule
\end{tabularx}
\end{table}

\textbf{With \rqtwoactionpast tool descriptions, agents achieve an absolute increase of 5.85 percentage points (median) in task success rate across all models and domains.}
As summarized in~\tabref{tab:optimized_sr_comparison}, we report the absolute success rate change ($\Delta$SR = SR$_{\text{after \rqtwonoun}}$ -- SR$_{\text{original}}$) for each model and domain to avoid inflation effects caused by low baseline success rates. Across four foundation models and twenty-four benchmark runs, agents using \rqtwoactionpast descriptions outperform their baseline counterparts in 54.17\% of cases (highlighted in green). While we have observed that the tool description scores have improved for all components (in~\tabref{tab:wilcoxon_bo_ao_full}), the performance of the agent regresses in 16.67\% of cases (highlighted in red), indicating that \rqtwoaction all components does not necessarily improve the success rate in all domains and models. We observe the greatest improvement in the \textit{Finance} domain for the \model{Qwen3-Coder-480B-A35B} model, whereas a moderate decline occurs for \model{GLM-4.5} in the \textit{Web Searching} domain. Despite these regressions, McNemar’s test confirms that the 5.85 percentage point improvement in SR is statistically significant (with $p = 0.02$) across the full benchmark. These results suggest that \rqtwoaction tool descriptions across all components can substantially improve task success rates in many domains, while also emphasizing the need for adaptive \rqtwonoun strategies tailored to specific domain contexts.

\begin{table}[t]
\centering
\small
\caption{Comparison of baseline (Base.) and \rqtwoactionpast (Aug.) results for success rate (SR), average evaluator score (AE), and average number of steps (AS) across models. Baseline SR, AE, and AS values for \model{GPT-4.1}, \model{Qwen3-Coder-480B-A35B}, and \model{GLM-4.5} are taken from the MCP-Universe baseline study~\cite{luo2025mcp}, whereas the baseline values for \model{Qwen3-Next-80B-A3B-Instruct} are obtained by running the agent with the original tool descriptions. Here, SR is aggregated over the models from~\tabref{tab:optimized_sr_comparison}. Column ``\# Tasks AE$\geq$0.80'' indicates the number of tasks that passed 80\% of the evaluators. The last row reports the overall median change; green marks improvement in SR or AE, and red marks degradation in AS as a trade-off for performance gain.}
\label{tab:overall_metrics_comparison}
\begin{tabularx}{\textwidth}{
  >{\raggedright\arraybackslash}p{2.6cm}   
  >{\raggedleft\arraybackslash}p{1.2cm}    
  >{\raggedleft\arraybackslash}p{1.2cm}    
  >{\raggedleft\arraybackslash}p{1.2cm}    
  >{\raggedleft\arraybackslash}p{1.2cm}    
  >{\raggedleft\arraybackslash}p{1.4cm}    
  >{\raggedleft\arraybackslash}p{1.2cm}    
  >{\raggedleft\arraybackslash}p{1.2cm}    
}
\toprule
\textbf{Model name}
& \multicolumn{2}{c}{\textbf{Overall SR}}
& \multicolumn{3}{c}{\textbf{Overall AE}}
& \multicolumn{2}{c}{\textbf{Overall AS}} \\
\cmidrule(lr){2-3} \cmidrule(lr){4-6} \cmidrule(lr){7-8}
 & \textbf{Base.} & \textbf{Aug.}
 & \textbf{Base.} & \textbf{Aug.} & \textbf{\# Tasks AE$\geq$0.80}
 & \textbf{Base.} & \textbf{Aug.} \\
\midrule
\model{GPT-4.1}                         & 18.18 & 29.44 & 0.41 & 0.47 & 19 & 5.24 & 8.08 \\
\model{Qwen3-Coder-480B-A35B }          & 19.91 & 25.97 & 0.38 & 0.43 & 16 & 7.78 & 14.06 \\
\model{GLM-4.5}                         & 24.68 & 25.25 & 0.41 & 0.45 & 18 & 7.33 & 14.79 \\
\model{Qwen3-Next-80B-A3B-Instruct}     & 15.58 & 21.21 & 0.33 & 0.39 & 16 & 9.46 & 6.97 \\
\midrule
\textbf{Median change}         &       & \cellcolor{green!25}\textbf{5.85}
                                &       & \cellcolor{green!25}\textbf{15.12}
                                & 17
                                &       & \cellcolor{red!20}\textbf{67.46} \\
\bottomrule
\end{tabularx}
\end{table}

\textbf{Beyond task-level success rate, \rqtwoactionpast tool descriptions also improve evaluator-level performance, increasing the Average Evaluator Score (AE) by 15.12\% across all foundation models}. As shown in~\tabref{tab:overall_metrics_comparison}, the overall AE, which represents the mean proportion of evaluators passing per task (where AE = 1 denotes complete success) increased consistently across all four models. This indicates that even for tasks not fully completed (i.e., not satisfying all evaluators), \rqtwoactionpast descriptions help agents satisfy a greater share of evaluation criteria. Using the Wilcoxon signed-rank procedure, we determine that the observed increase in AE is statistically significant, with $p<0.01$. Furthermore, we observe that 17 tasks, or 7.36\% of all tasks (median across models), can achieve an AE $\geq 0.80$ (i.e., passing more than 80\% of evaluators) yet still fall short of final success, often because the benchmark imposes a maximum iteration limit. These trends suggest that \rqtwonoun enhances intermediate reasoning and partial goal completion, aligning with prior findings~\cite{sahoo2024systematic, zheng2023progressive} that improved prompt clarity strengthens reasoning pathways to drive steadier, more progressive task execution.

\textbf{The average number of execution steps (AS) increases by 67.46\% (median) across models and domains when using the fully \rqtwoactionpast tool descriptions, compared to the original MCP-Universe baseline}. As shown in~\tabref{tab:overall_metrics_comparison}, despite the absolute AS remaining relatively low (less than 15 for two models and 9 for two others), three of the four evaluated models show a statistically significant increase in AS (with $p<0.001$ at the Wilcoxon signed-rank test). In contrast, the smaller-sized \model{Qwen3-Next-80B-A3B-Instruct} model shows resilience, reducing AS from the baseline while still improving both Success Rate (SR) and Average Evaluator Score (AE). 

\textbf{Similarly, the use of augmented tool descriptions inherently increases the token consumption of agents}. As shown in~\tabref{tab:token_usage_comparison}, we measure median input and output token usage before and after augmentation for \model{Qwen3-Next-80B-A3B-Instruct} across three MCP-Universe domains: Finance, Repository Management, and 3D Design. Because the original MCP-Universe benchmark does not report token-level statistics, we conduct this token analysis using \model{Qwen3-Next-80B-A3B-Instruct}. We choose \model{Qwen3-Next-80B-A3B-Instruct} to guarantee methodological consistency, as we have already employed it to derive other fine-grained metrics absent from the original study, namely, task-level success rate, AE, and AS. Evaluating both the original and augmented descriptions under this unified setup ensures that all supplementary measurements are drawn from the exact same model and execution environment.  Also, we sample Finance, Repository Management, and 3D Design as these domains do not require additional, expensive service API keys like other domains (such as Google Maps or Search API keys for Location Navigation or Web Search Domain). Across these domains, augmented descriptions increase median input tokens by 21.85–41.43\% and median output tokens by 11.10–69.00\%. These results confirm that richer descriptions improve performance at the cost of higher token usage.

For the rest of the analysis, we use Average Steps (AS) as the primary cost proxy for two reasons. First, AS is reported by MCP-Universe and is therefore available for comparison across all evaluated models and domains, whereas token-level statistics are not reported for the original benchmark runs. Second, our sampled token analysis is directionally consistent with AS: augmented descriptions increase token usage and also increase execution steps. Therefore, AS provides a benchmark-compatible proxy for execution cost, while the token analysis confirms that this proxy accurately mirrors the underlying trends in actual token expenditure.

\begin{figure}[t]
    \centering

    \begin{subfigure}[b]{0.45\linewidth}
        \centering
        \includegraphics[width=\linewidth]{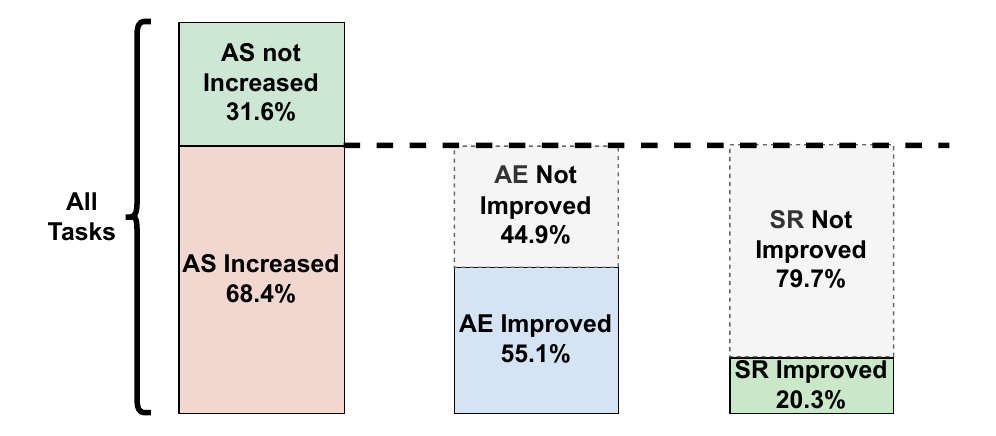}
        \caption{\model{GPT-4.1}}
    \end{subfigure}
    \hfill
    \begin{subfigure}[b]{0.45\linewidth}
        \centering
        \includegraphics[width=\linewidth]{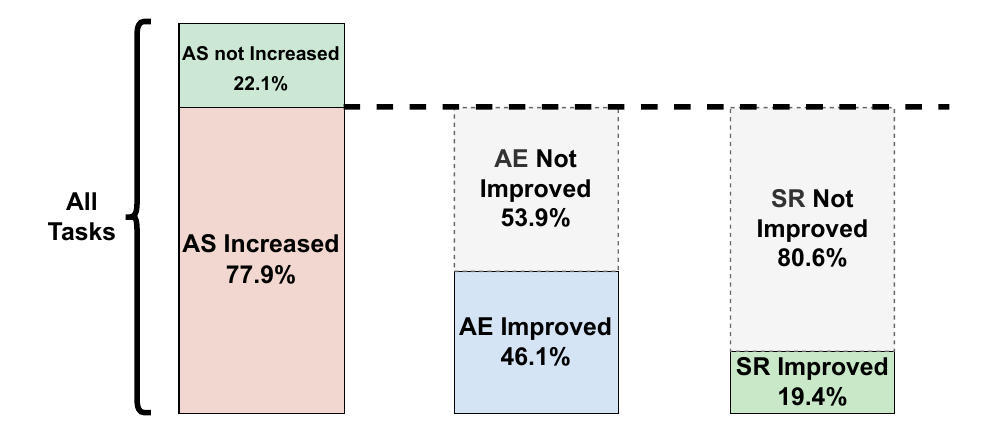}
        \caption{\model{GLM-4.5}}
    \end{subfigure}

    \vspace{1em}

    \begin{subfigure}[b]{0.45\linewidth}
        \centering
        \includegraphics[width=\linewidth]{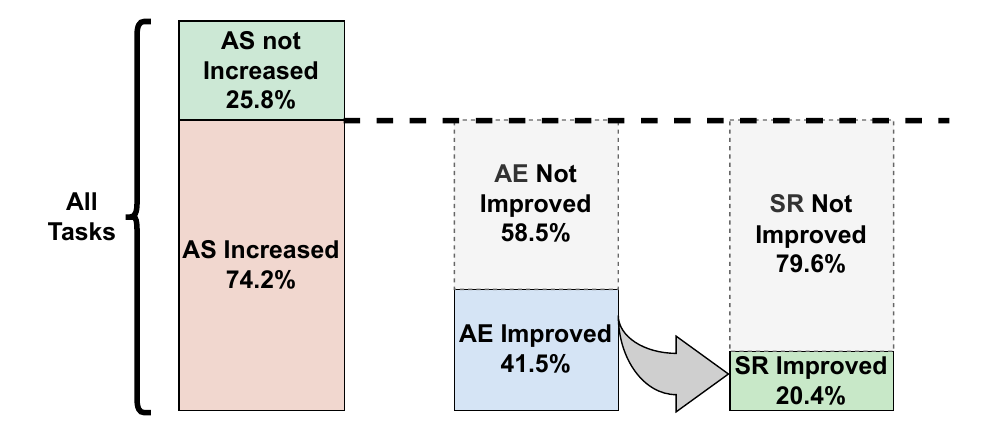}
        \caption{\model{Qwen3-Next-80BA3B-Instruct}}
    \end{subfigure}

    \caption{Task flow from increased steps (AS) to improved evaluator completion (AE) and final success (SR). For each model, the leftmost bar shows all tasks split by whether AS increased from the baseline. The horizontal dotted line indicates that only the AS-increased subset (shown in red in the lower segment of the leftmost bar) is considered in the subsequent bars. Accordingly, the AE and SR bars report the fractions of AS-increased tasks that (i) improve in AE and (ii) improve in SR, respectively.}
    \label{fig:as-ae-sr-conversion}
\end{figure}

However, focusing on the broader trend independent of domain (Figure~\ref{fig:as-ae-sr-conversion}), 68--78\% of tasks require more steps than the baseline. Among these tasks with increased AS, roughly half (41--55\%) show improved AE; however, only 19--20\% of all tasks achieve the final success. This funnel indicates that richer descriptions motivate deeper intermediate exploration that captures more requirements, even if it does not always yield a perfect final output. For instance, in a \textit{Location Navigation} task (google\_maps\_task\_0001), the fully \rqtwoactionpast tool description doubled the agent's steps (from 9 to 19) compared to the baseline with \model{Qwen3-Next-80B-A3B-Instruct}. However, it increased the number of passing evaluators from 8 to 10 out of 11. Consequently, the aggregated increase in AS represents a cost-performance trade-off: where agents expend more computational effort to achieve greater partial progress and a higher likelihood of final completion.


\begin{table}[t]
\centering
\small
\caption{Median input and output token usage, in thousands (K), before (BA) and after (AA) tool description augmentation across selected MCP-Universe domains for \model{Qwen3-Next-80B-A3B-Instruct}. Augmented descriptions increase both input and output tokens, reflecting the token-cost side of the performance trade-off.}
\label{tab:token_usage_comparison}
\begin{tabularx}{\textwidth}{
  >{\raggedright\arraybackslash}p{3cm}   
  >{\raggedright\arraybackslash}p{2cm}   
  >{\raggedleft\arraybackslash}p{1.0cm}  
  >{\raggedleft\arraybackslash}p{1.0cm}  
  >{\raggedleft\arraybackslash}p{1.0cm}  
  >{\raggedleft\arraybackslash}p{1.0cm}  
  >{\raggedleft\arraybackslash}p{1.0cm}  
  >{\raggedleft\arraybackslash}p{1.0cm}  
}
\toprule
\textbf{Model} & \textbf{Domain} 
& \multicolumn{3}{c}{\textbf{Input Tokens (K)}} 
& \multicolumn{3}{c}{\textbf{Output Tokens (K)}} \\
\cmidrule(lr){3-5} \cmidrule(lr){6-8}
 & 
 & \textbf{BA} & \textbf{AA} & \textbf{$\Delta$\%} 
 & \textbf{BA} & \textbf{AA} & \textbf{$\Delta$\%} \\
\midrule
\multirow{3}{=}{\model{Qwen3-Next-80B-A3B-Instruct}}
& Finance          & 50.64  & 61.71  & 21.85 & 0.52 & 0.68 & 30.35 \\
& Repo Management  & 232.41 & 299.86 & 29.02 & 2.84 & 3.15 & 11.10 \\
& 3D Design        & 43.19  & 61.09  & 41.43 & 2.74 & 4.63 & 69.00 \\
\bottomrule
\end{tabularx}
\end{table}

\begin{figure*}[t] 
  \centering
  
  \begin{subfigure}[t]{0.32\textwidth}
    \centering
    \includegraphics[width=\linewidth]{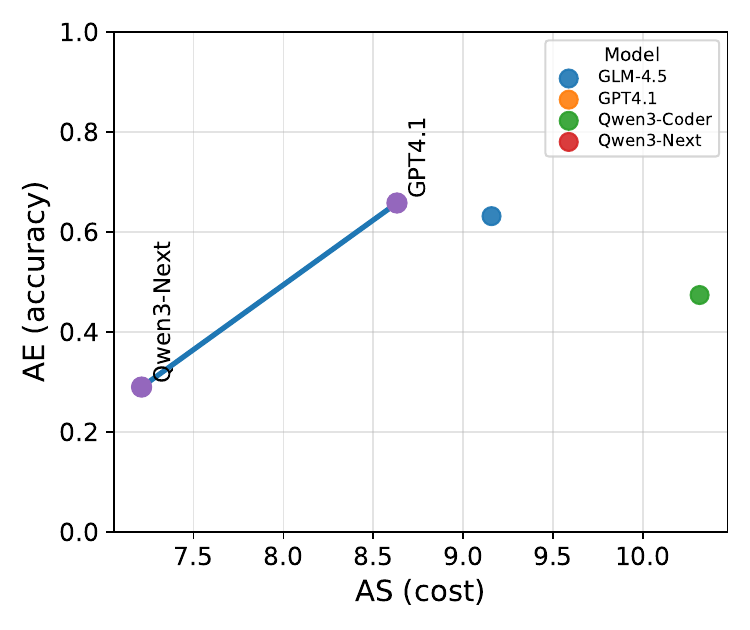}
    \caption{3D design}
    \label{fig:pareto:3d_design}
  \end{subfigure}\hfill
  \begin{subfigure}[t]{0.32\textwidth}
    \centering
    \includegraphics[width=\linewidth]{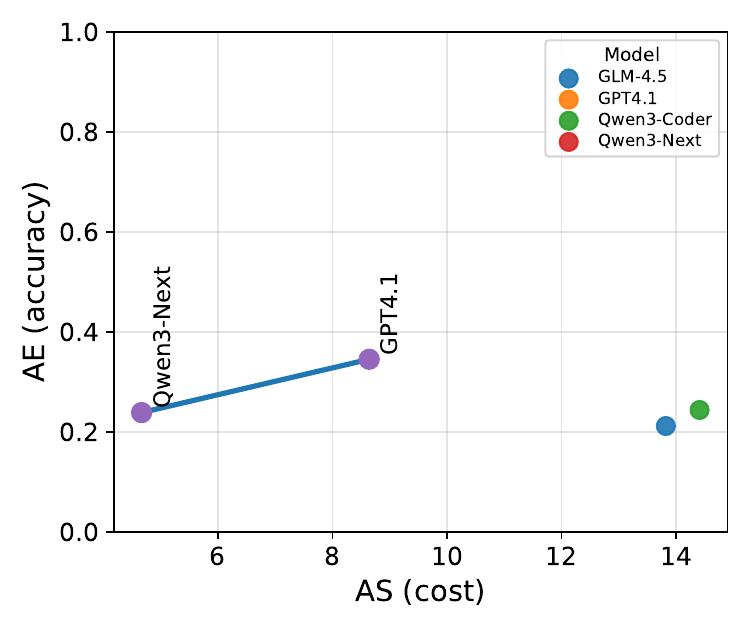}
    \caption{Browser automation}
    \label{fig:pareto:browser_automation}
  \end{subfigure}\hfill
  \begin{subfigure}[t]{0.32\textwidth}
    \centering
    \includegraphics[width=\linewidth]{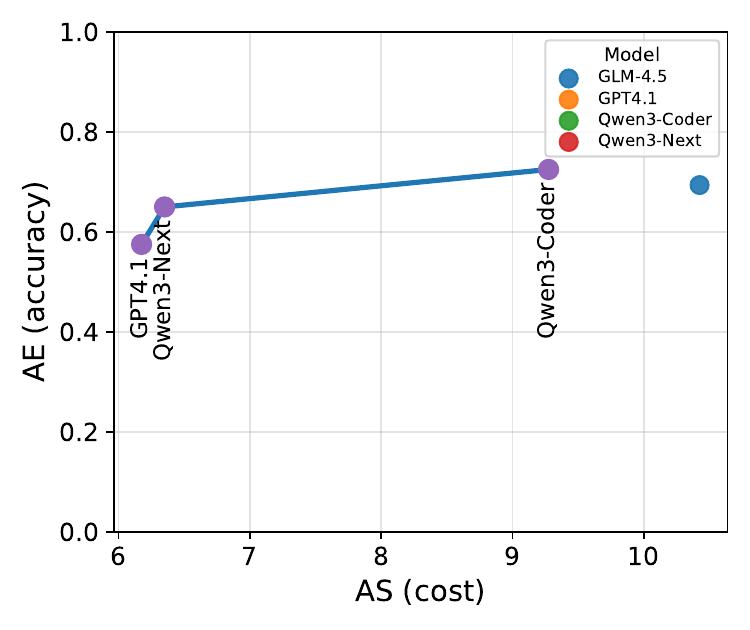}
    \caption{Financial analysis}
    \label{fig:pareto:financial_analysis}
  \end{subfigure}

  \vspace{0.5em} 

  \begin{subfigure}[t]{0.32\textwidth}
    \centering
    \includegraphics[width=\linewidth]{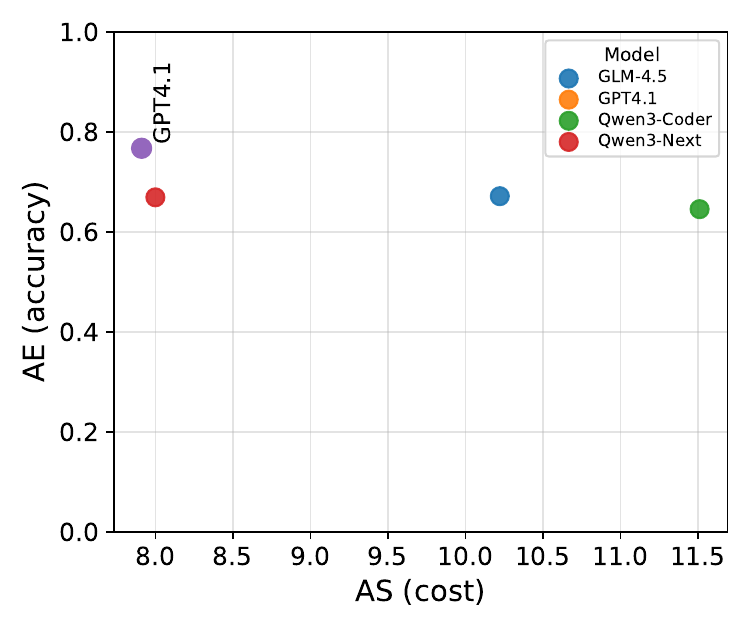}
    \caption{Location navigation}
    \label{fig:pareto:location_navigation}
  \end{subfigure}\hfill
  \begin{subfigure}[t]{0.32\textwidth}
    \centering
    \includegraphics[width=\linewidth]{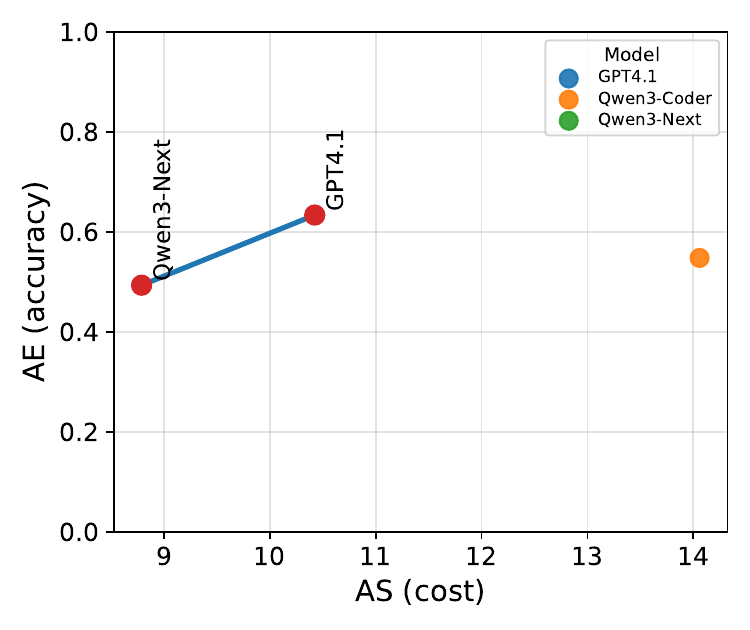}
    \caption{Repository management}
    \label{fig:pareto:repository_management}
  \end{subfigure}\hfill
  \begin{subfigure}[t]{0.32\textwidth}
    \centering
    \includegraphics[width=\linewidth]{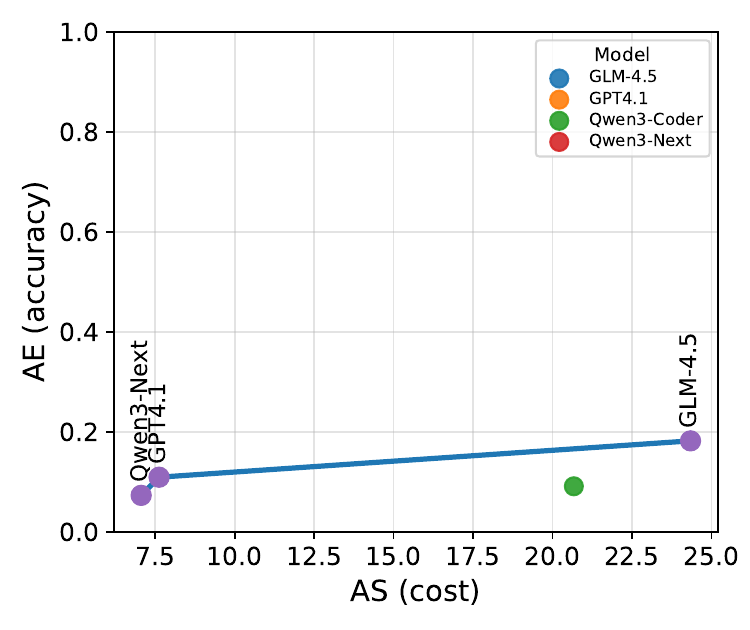}
    \caption{Web search}
    \label{fig:pareto:web_search}
  \end{subfigure}
  \caption{Domain-wise Pareto frontiers. Each subfigure shows the trade-off between Average Evaluator Score (AE; proxy for accuracy) and Average Steps (AS; proxy for cost) for different models within a domain. All subfigures share an identical AE axis for comparability, while the AS axis is scaled independently to reflect domain-specific execution costs.}
  \label{fig:pareto:all_domains}
\end{figure*}

\textbf{While the proprietary \model{GPT-4.1} leads in peak performance among the evaluated models, a relatively smaller-sized open-weight model, \model{Qwen3-Next-80B-A3B-Instruct}, demonstrates a superior balance between cost and accuracy compared to substantially larger open-weight alternatives under \rqtwoactionpast tool descriptions}. As shown in Figure~\ref{fig:pareto:all_domains}, the Pareto frontiers visualize the trade-off between Average Evaluator Score (AE) and Average Steps (AS) across six domains when agents operate with \rqtwoactionpast tool descriptions. Models falling \textit{on} the blue line represent efficient choices. Models falling \textit{below} the line (e.g., \model{Qwen3-Coder-480B} in \textit{3D Design}) are suboptimal, as they consume more steps to achieve equal or lower accuracy. Across domains such as \texttt{Browser Automation}, \texttt{Financial Analysis}, \texttt{Location Navigation}, and \texttt{Repository Management}, \model{Qwen3-Next-80B-A3B-Instruct} consistently appears on or near the Pareto frontier, achieving competitive AE while maintaining relatively low AS. At the same time, \model{GPT-4.1} frequently lies on the Pareto frontier and is often closest to the idealized upper-left region, reflecting strong accuracy with moderate execution cost. In contrast, larger open-weight models like \model{GLM-4.5} and \model{Qwen3-Coder-480B-A35B} generally occupy dominated regions with higher AS for comparable AE. These observations indicate that under \rqtwoactionpast tool descriptions, a simple parameter scale does not guarantee higher efficiency; rather, different architectures occupy distinct niches in the cost-accuracy trade-off space.

\begin{Summary}{Summary of RQ-2}{
\begin{enumerate}
    \item Augmenting tool descriptions with all components can improve agents' performance, increasing the Average Evaluator Score by \textbf{15.12\%} and the task success rate by \textbf{5.85 percentage points}, although \textbf{16.67\%} of cases experience regressions.
    \item As a trade-off for these gains, the median number of execution steps increases by \textbf{67.46\%}, reflecting deeper exploration to solve problems.
    \item Model scale alone does not guarantee efficiency; smaller-sized open-weight models, such as \model{Qwen3-Next-80B-A3B-Instruct}, achieve a superior cost-accuracy balance compared to larger open-weight models, like \model{Qwen3-Coder-480B-A35B} or \model{GLM-4.5}.
\end{enumerate}
}
\end{Summary}

\subsection{RQ-3: How do different components of the \rqtwoactionpast tool description impact the performance of FM-based agents?}\label{results-rq3}

\motivation
Although fully \rqtwoactionpast tool descriptions can improve agent performance, it remains unclear which description components drive these gains and which ones are redundant or even detrimental. Practitioner guidance is also inconsistent; for instance, Anthropic considers the \textit{Examples} component less critical~\cite{anthropic-tool-description}, whereas other studies suggest it is beneficial~\cite{xu2025llm}. Furthermore, given that fully \rqtwoactionpast tool descriptions can overload the FM's context window and conflict with the increasing adoption of progressive capability disclosure via agent skills, identifying a minimal effective set of description components is essential to preserve performance while reducing overhead. To this end, we conduct a systematic component ablation study~\cite{hameed2022based}, analyzing interactions among models, domains, and component configurations across both high- and low-performing settings.

\approach
We evaluate the impact of individual components of the \rqtwoactionpast tool description by conducting an ablation study across five (model, domain) combinations. In the RQ-2 results, we observe that with a fully \rqtwoactionpast tool description, i.e., one that includes all components, the SR of the agent can be improved for certain domain-model combinations, while potentially experiencing some regression in other domain-model combinations. For example, in the Finance domain, we have observed that with a fully \rqtwoactionpast tool description, all models have shown higher SR. Conversely, for 3D Design, with an \rqtwoactionpast tool description, \model{GPT-4.1} has achieved higher SR, whereas \model{Qwen3-Coder-480B-A35B} has shown lower SR than the baseline. To investigate the mechanics behind these divergent outcomes, we select domain-model combinations from Table~\ref{tab:optimized_sr_comparison} that cover both performance profiles. We specifically examine three combinations where full \rqtwonoun yields gains: \textit{Finance} with \model{GPT-4.1}, \textit{Location Navigation} with \model{GPT-4.1}, and \textit{Repository Management} with \model{Qwen3-Coder-480B-A35B}. Conversely, to understand regression modes, we also include two combinations where performance degrades relative to the baseline: \textit{Web Searching} with \model{GLM-4.5} and \textit{3D Design} with \model{Qwen3-Coder-480B-A35B}.

For each combination, we run two ablation variants: (i) removing the \textit{Examples} component from all tool descriptions, and (ii) combining two components where one is always \textit{Purpose}. We keep \textit{Purpose} fixed in all combinations because it defines the core functionality of the tool, which is essential for agents to understand what the tool does and when to use it. We execute the agents under these settings while controlling the composition of the descriptions through the \texttt{--components} flag introduced in~\secref{mcp-client-calibration}. The results from these controlled runs allow us to isolate the relative influence of specific components on agent performance across diverse model and domain contexts. To validate whether various ablation combinations exhibit statistically significant behavioral consistency, we employ Pearson’s Chi-Squared test of independence~\cite{zibran2007chi} complemented by the $\phi$-signed coefficient, which quantifies the strength of association between the two configurations.

\findings
\textbf{There is no single ``golden'' combination of components that yields the best results across all domain-model pairs}. As summarized in~\tabref{tab:component_wise_ablation_study}, the best-performing component combination varies across domain-model pairs. For instance, in the Finance domain with \model{GPT-4.1}, using only the \textit{Purpose} and \textit{Guidelines} components yields the highest success rate, surpassing the fully \rqtwoactionpast tool description, making it the best-performing component combination (BC). In contrast, tasks in Location Navigation with the same model show the best performance with the fully \rqtwoactionpast tool description (FR). Meanwhile, in Repository Management with \model{Qwen3-Coder-480B-A35B}, performance improves modestly when only the \textit{Examples} component is removed. These results reveals a complex interplay where the effectiveness of specific rubric components varies across settings, depending on both the underlying model architecture and the domain-specific requirements.

To analyze why the combination of \textit{Purpose} and \textit{Guideline} is improving the SR for the Finance domain, we examine the tool descriptions further. We observe that in the tool \texttt{get\_historical\_stock\_prices} in the \texttt{yfinance} MCP server, the \textit{Guidelines} component provides critical operational cues such as ``requested dates should include trading days'' and ``set \texttt{end\_date} one day later than expected since the tool returns the previous day's closing price.'' These explicit behavioral instructions help the model reason correctly about valid input ranges and temporal offsets, leading to higher task success when this component is used alone. In contrast, the \textit{Limitations} component of the same tool includes vague or self-referential statements such as ``this contradiction requires disambiguation before relying on intraday availability'', which can introduce uncertainty into the model's reasoning. When combined with other components, such ambiguity dilutes otherwise useful guidance and lowers performance relative to the single-component configuration. This pattern suggests that components that convey precise behavioral constraints of a tool improve agent performance, whereas components containing ambiguous or contradictory statements can degrade it.

\begin{table}[t]
\centering
\small
\caption{Success rate (SR) comparison across component combinations (rows) for each domain-model pair (columns). Green marked ones are the highest performance achieved in the respective domain-model combination.
\textbf{Notation:} FR = Fully \rqtwoactionpast tool description containing all components; P = Purpose; G = Guidelines; L = Limitation; PEx = Parameter Explanation; E = Examples.}
\label{tab:component_wise_ablation_study}
\begin{tabularx}{\textwidth}{
  >{\raggedright\arraybackslash}p{2.6cm}  
  *{6}{>{\raggedleft\arraybackslash}X}    
}
\toprule
\textbf{Rubric setup} &
\textbf{Finance (\model{GPT-4.1})} &
\textbf{Location (\model{GPT-4.1})} &
\textbf{Repo (\model{Qwen3-Coder})} &
\textbf{3D-design (\model{Qwen3-Coder})} &
\textbf{Web Searching (\model{GLM-4.5})} &
\textbf{Median SR}\\
\midrule
\textbf{FR}          & 57.50\% & \cellcolor{green!25}\textbf{31.00\%} & 18.20\% & 21.10\% & \cellcolor{green!25}\textbf{18.18}\% & 21.10\% \\
\textbf{P + G + L + PEx} & 55.00\% & 26.70\% & \cellcolor{green!25}\textbf{21.21\%} & 21.10\% & 12.73\% & 21.21\%\\
\textbf{P + G}       & \cellcolor{green!25}\textbf{67.50\%} & 20.00\% & 18.20\% & 15.80\% & 12.73\% & 18.20\% \\
\textbf{P + L}       & 47.50\% & 24.50\% & 18.20\% & 21.05\% & 16.36\% & 21.05\%\\
\textbf{P + E}       & 62.50\% & 17.80\% & 18.20\% & \cellcolor{green!25} \textbf{26.32\%} & 10.91\% & 18.20\%\\
\textbf{P + PEx}     & 40.00\% & 20.00\% & 6.06\%  &15.80\% &\cellcolor{green!25}\textbf{18.18\%} & 18.18\%\\
\bottomrule
\end{tabularx}
\end{table}

\textbf{Statistical analysis shows a strong association between task-level outcomes produced by the fully \rqtwoactionpast tool description (FR) and those produced by the best-performing component combination (BC).} As shown in~\tabref{tab:ablation-chi-squared-test}, Pearson’s chi-square tests reveal significant dependencies between BC and FR across multiple domain-model combinations, including \texttt{3D Design}, \texttt{Financial Analysis}, and \texttt{Repository Management} (\textit{p} < 0.01). These results indicate that task success under BC and FR is associated rather than independent, meaning that both configurations tend to succeed or fail on the same tasks. The $\phi$-signed coefficient, which measures agreement between paired binary outcomes, ranges approximately from 0.5 to 0.9, indicating a strong correspondence in the sets of tasks solved by BC and FR. This correspondence suggests that, for a given domain-model combination, BC preserves much of the functional guidance encoded in FR, despite omitting certain components. Consequently, domain-specific pruning emerges as a favorable engineering trade-off that can maintain logical reliability comparable to the fully \rqtwoactionpast description, while reducing token usage and inference latency, provided that the pruning strategy is tailored to the target domain.

\begin{table}[t]
\centering
\small
\caption{Comparison between the fully \rqtwoactionpast tool description containing all components (FR) and the best-performing component combination (BC). 
Panel~A reports Pearson's chi-square tests assessing whether task success under FR and BC is statistically dependent. Higher signed $\phi$ values indicate stronger agreement between paired binary outcomes, implying that BC and FR solve similar tasks. 
Panel~B shows the overall confusion matrix of task outcomes between the two configurations, where a value of 1 indicates tasks solved and 0 indicates tasks failed.}
\label{tab:ablation-chi-squared-test}
\begin{tabularx}{\textwidth}{
  >{\raggedright\arraybackslash}p{3.2cm}
  >{\raggedright\arraybackslash}p{2.4cm}
  >{\raggedleft\arraybackslash}p{2.2cm}
  >{\raggedleft\arraybackslash}p{2.2cm}
}
\toprule
\multicolumn{4}{l}{\textbf{Panel A: Pearson's chi-square tests}} \\
\midrule
\textbf{Domain} & \textbf{Model} & \textbf{p-value} & \textbf{$\phi$ (signed)} \\
\midrule
3D Design             & \model{Qwen3-Coder} & $<0.01$ & 0.864 \\
Financial Analysis    & \model{GPT-4.1}     & $<0.01$ & 0.591 \\
Location Navigation   & \model{GPT-4.1}     & $<0.01$ & 0.572 \\
Repository Management & \model{Qwen3-Coder} & $<0.01$ & 0.909 \\
Web Searching            & \model{GLM-4.5}     & $<0.01$ & 0.511 \\
\midrule
\multicolumn{4}{l}{\textbf{Panel B: Overall confusion matrix (all domains)}} \\
\textbf{} & \textbf{FR=1} & \textbf{FR=0} &  \\
\textbf{BC=1} & 46 & 15 &  \\
\textbf{BC=0} & 11 & 120 &  \\
\bottomrule
\end{tabularx}
\end{table}

\textbf{Removing the \textit{Examples} component does not significantly degrade performance compared to either the best-performing component combination (BC) or the fully \rqtwoactionpast tool description (FR), contradicting traditional benefits of few-shot examples.} As shown in~\tabref{tab:component_wise_ablation_study}, configurations excluding \textit{Examples}, such as \textit{P+G+L+PEx}, achieve stable success rates across approximately 60\% of the evaluated domain-model pairs, although they are not always the highest-performing configuration. Furthermore, the Median SR confirms this stability: \textit{P+G+L+PEx} achieves the highest median SR of 21.21\% across all domain-model pairs, marginally surpassing FR (21.10\%), further supporting that dropping \textit{Examples} does not hurt and may even slightly improve cross-domain consistency. To jointly assess performance differences across the three configurations (FR, BC, and \textit{P+G+L+PEx}), we apply Cochran’s Q test~\cite{cohen2015cochran}, which is appropriate for comparing three or more matched binary outcomes. Across all evaluated domain-model combinations, Cochran’s Q test consistently yields $p>0.20$, indicating no statistically significant differences among the three configurations. This result suggests that the inclusion or removal of the \textit{Examples} component does not materially affect task success rates, which affirms Anthropic's suggestion to put less emphasis on examples, but contradicts the traditional benefit of few-shot examples in the prompts~\cite{brown2020language} of MCP tool descriptions.

\begin{figure*}[t]
	\centering
	\includegraphics[width=0.99\textwidth]{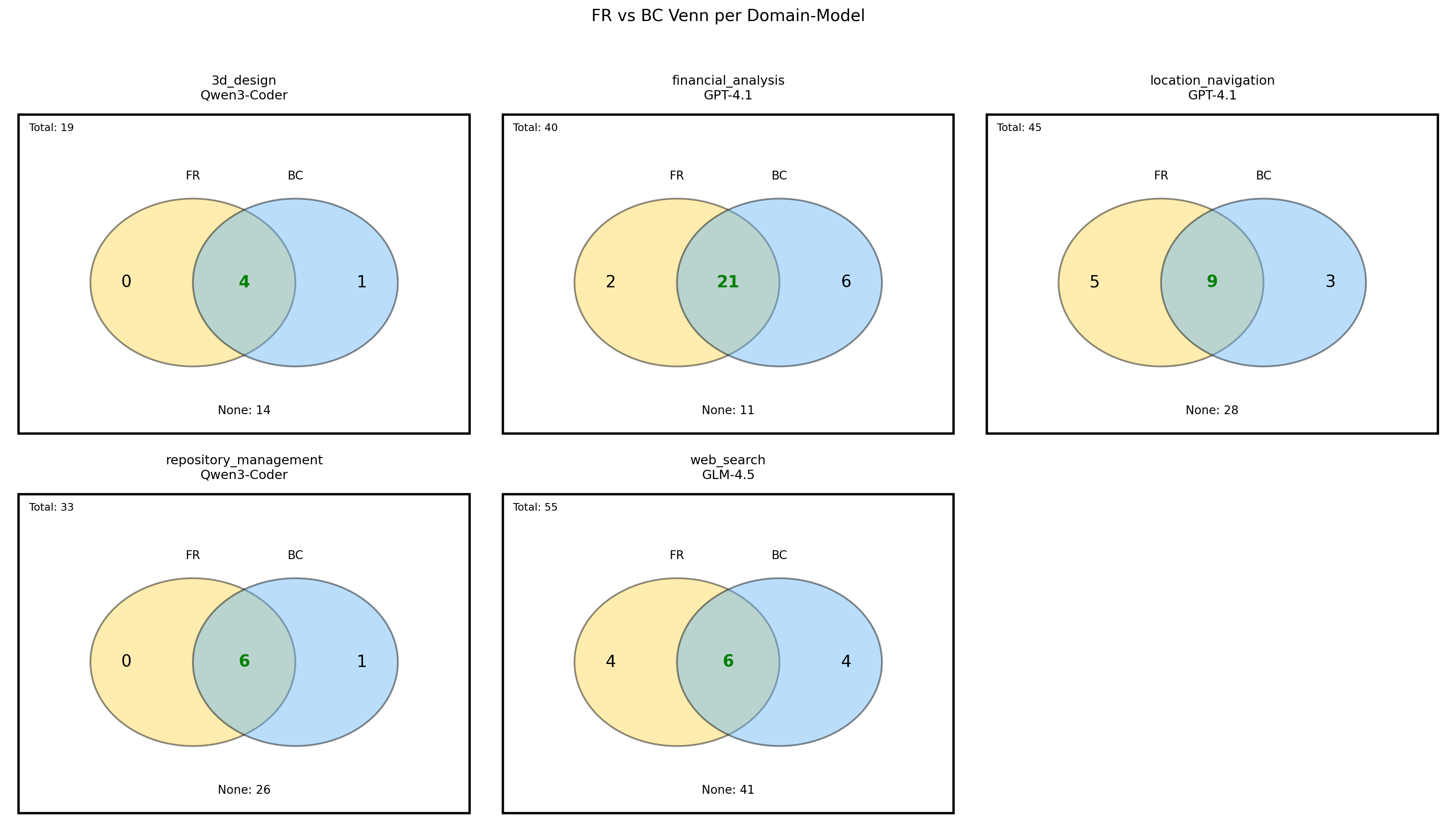}
	\caption{Overlap between tasks solved by the fully \rqtwoactionpast tool description (FR) and the best-performing reduced component configuration (BC) across five domain-model combinations. FR (yellow) denotes tasks successfully solved using the fully \rqtwoactionpast tool description, BC (blue) represents tasks solved using the best-performing reduced component configuration, and Green intersection indicates tasks solved by both configurations.}
	\label{ablation:venn-diagram}
\end{figure*}

\textbf{The best-performing component combinations (BC) and their fully \rqtwoactionpast counterparts (FR) solve overlapping but distinct sets of tasks, indicating complementary coverage across configurations}. As illustrated in Figure~\ref{ablation:venn-diagram}, neither BC nor FR fully subsumes the task coverage of the other. Instead, each solves a partially unique subset of tasks that have a high overlap. For example, in \texttt{Location Navigation} with \model{GPT-4.1}, the two configurations share nine successfully solved tasks but diverge on eight others, suggesting that different component structures guide the model toward distinct reasoning trajectories. Hypothetically, when the results from both configurations are combined in a hybrid approach, the overall success rate has the potential to increase; for instance, running the failed \texttt{Location Navigation} tasks with the reduced components after the fully \rqtwoactionpast one can raise the success rate to 37.8\%. While such an ensemble approach inherently increases cost (in terms of steps and latency), it opens a viable research direction for maximizing accuracy in mission-critical scenarios where a high success rate is more important than cost.

\begin{Summary}{Summary of RQ-3}{
\begin{enumerate}
    \item There is no single ``golden'' combination of tool description components; effectiveness is highly context-dependent across different models and domains. 
    \item Removing the \textit{Examples} component does not result in statistically significant performance degradation ($p>0.20$), supporting Anthropic’s guidance on deprioritizing examples while contrasting with the traditional benefits attributed to few-shot prompting.
    \item Depending on the domain and model, tool descriptions can be pruned to a small set of components whose task-level outcomes are strongly associated with those of fully \rqtwoactionpast descriptions, offering a cost-efficient alternative without sacrificing logical reliability.
\end{enumerate}
}
\end{Summary}
\section{Discussion}
\subsection{Impact of Source Code Grounding in Smell Detection}\label{source-code-grounded-eval}
\textbf{Source-code evidence can help FM evaluators assess whether a tool description is consistent with the tool’s actual implementation}. In the default rubric-based evaluation, the LLM-as-Jury receives the original tool description, input schema, and scoring rubric. However, some descriptions may be outdated, incomplete, or inconsistent with the implementation. In such cases, evaluating only the description and schema may not be sufficient to assess whether missing details, parameter explanations, or stated constraints are faithful to the tool’s behavior. To reduce this risk, we conduct an additional source-code-grounded evaluation for tools whose implementation is publicly available.

For each eligible MCP tool, we statically identify and extract the corresponding implementation slice from the MCP server repository. We focus on Python, JavaScript/TypeScript, and Go repositories that follow common MCP SDK, FastMCP-style, or \texttt{mcp-go} registration patterns. For Python tools, we use the \texttt{ast} module to identify MCP-style tool decorators, such as \texttt{@tool} and \texttt{@mcp.tool}. For JavaScript and TypeScript tools, we detect common registration patterns, such as \texttt{.tool(...)} and \texttt{server.setRequestHandler(...)}, and resolve nearby schema definitions used to declare tool inputs. For Go tools, we identify \texttt{mcp.NewTool(...)} registrations and recover parameters declared through \texttt{mcp.With*} option builders, including simple helper functions that return tool options. We also support OpenAPI Spec~\cite{dos2020analysis} parsing to identify grounding sources for API-based MCP tools. When explicit registration patterns are not found, we fall back to text matching using the tool name and original description. We then provide the extracted implementation slice to the FM evaluator as an explicit grounding block, together with the original description, input schema, and rubric.

To quantify the impact of source-code grounding on rubric-based evaluation, we identify source code for 183 of the 202 MCP-Universe tools from public repositories and the OpenAPI spec for 17 of them.  The source code for the remaining 2 tools was not publicly available. We then score each tool-description component twice using \model{gpt-4.1-mini}: once without source-code grounding and once with the extracted source-code slice included as grounding evidence. We compare the paired scores using the Wilcoxon signed-rank test to determine whether grounding changes the evaluator’s rubric judgments. We report matched-pairs rank-biserial correlation ($r$) as the Wilcoxon effect size, ranging from $-1$ to $+1$. Following  Maciej et al.~\citep{tomczak2014need}, we interpret $|r|$ values of 0.10--0.30 as small, 0.30--0.50 as medium, and $\geq 0.50$ as large.

\begin{table}[t]
\centering
\small
\caption{Wilcoxon signed-rank comparison of rubric-component scores for the tools of MCP-Universe, evaluated without source-code grounding (UG) and with source-code grounding (WG). Each component is scored on a 1-5 ordinal scale, where higher values indicate better description quality. We report the Wilcoxon statistic, $p$-value, and the median score per condition. The rank-biserial effect size r is reported only for components with a significant difference (N/A otherwise).}
\label{tab:grounding_wilcoxon_rubric_scores}
\begin{tabular}{l r r r r r}
\toprule
\textbf{Component} & \textbf{Statistic} & \textbf{$p$-value} & \textbf{Med. Score (UG)} & \textbf{Med.Score (WG)} & \textbf{$r$} \\
\midrule
Purpose                 & 1908.00 & 1.00 & 2.00 & 2.00 & N/A \\
Guidelines              & 1274.00   & 1.00    & 1.10 & 1.00 & N/A \\
Limitations             & 2122.00   & 0.02    & 1.00 & 1.00 & 0.14 \\
Parameter Explanation   & 12265.50 & $<0.001$ & 1.00 & 2.00 & 0.60 \\
Examples                & 1038.00    & 1.00    & 1.00 & 1.00 & N/A \\
Length and Completeness & 4087.00  & 0.01 & 1.00 & 1.30 & 0.18 \\
\bottomrule
\end{tabular}
\end{table}

As shown in~\tabref{tab:grounding_wilcoxon_rubric_scores}, source-code grounding changes the evaluator’s judgments and generally increases rubric scores for three of the six components. The clearest improvement appears for \textit{Parameter Explanation}, where the median score increases from 1.0 to 2.0, with a statistically significant difference ($p<0.001$) and the largest effect size ($r = 0.60$). This suggests that source-code grounding helps the evaluator recognize parameter details that are recoverable from the implementation but absent or unclear in the original description. Overall, these results indicate that implementation-level evidence can improve the reliability of FM-based rubric scoring, particularly for components tied directly to tool behavior and input semantics.

As we reran the rubric evaluation on all MCP-Universe tools with LLM-as-Jury and found that source-code grounding increased the median Parameter Explanation score from 1.0 to 2.0, we updated the results in~\tabref{tab:wilcoxon_bo_ao_full} of~\secref{results-rq2} to~\tabref{tab:wilcoxon_bo_ao_full_updated}.

\begin{table}[t]
\centering
\small
\caption{Wilcoxon signed-rank test results comparing tool description component scores before \rqtwonoun (BA) grounded with source code and after \rqtwonoun (AA), showing statistically significant median score increases across all components. Here we recalculate the median scores before \rqtwonoun with the whole LLM-as-Jury and updated~\tabref{tab:wilcoxon_bo_ao_full}.}
\label{tab:wilcoxon_bo_ao_full_updated}
\begin{tabular}{lrrrrr}
\toprule
\textbf{Component} &
\textbf{Statistic} &
\textbf{$p$-value} &
\textbf{Med. Score (BA)} &
\textbf{Med. Score (AA)} &
\textbf{$r$} \\
\midrule
Purpose                   & 0.00 & $<0.001$ & 2.00 & 5.00 & 1.00 \\
Guidelines           & 0.00 & $<0.001$ & 1.00 & 5.00 & 1.00 \\
Limitations                & 0.00 & $<0.001$ & 1.00 & 5.00 & 1.00 \\
Parameter Explanation     & 11.00 & $<0.001$ & 2.00 & 4.70 & 1.00 \\
Examples                  & 0.00 & $<0.001$ & 1.00 & 5.00 & 1.00 \\
Length and Completeness   & 0.00 & $<0.001$ & 1.30 & 5.00 & 1.00 \\
\bottomrule
\end{tabular}
\end{table}

\subsection{Impact of Source Code Grounding in Augmentation}
\textbf{FM-based tool description augmentation should be grounded to reduce hallucinated or unsupported details}. Because many original tool descriptions are underspecified, an FM-based augmentor may complete missing information through unsupported inference. This creates a risk that the augmented description becomes fluent and complete-looking, but factually inconsistent with the tool’s actual behavior. This risk is especially relevant for components generated purely from prompts, i.e., \textit{Purpose}, \textit{Guidelines}, and \textit{Parameter Explanation} in~\secref{initial-augmentation}, where the augmentor may infer behavior, constraints, or edge cases not present in the original implementation (source code) of the tool.

To mitigate this risk, we ground the augmentation process in implementation-level evidence by incorporating the tool's source code extracted in~\secref{source-code-grounded-eval}. We provide the extracted source-code slice to the FM as an explicit grounding context, accompanied by the original tool description and input schema, and instruct the FM to avoid inventing details that are not supported by the code, input schema, or original description.

\begin{table}[t]
\centering
\small
\caption{Wilcoxon signed-rank comparison of component-level faithfulness for augmented tool descriptions, evaluated without source-code grounding (UG) and with source-code grounding (WG), over publicly available MCP-Universe tools. Faithfulness is scored on a 0-1 scale, where higher values indicate stronger support from the tool's source code and input schema. We report the Wilcoxon statistic, p-value, and the median faithfulness score per condition. The rank-biserial effect size r is reported for all components; positive values indicate that WG yields higher faithfulness than UG.}
\label{tab:faithfulness_ug_vs_wg}
\begin{tabular}{l r r r r r}
\toprule
\textbf{Component} & \textbf{Statistic} & \textbf{$p$-value} & \textbf{Med. Score (UG)} & \textbf{Med. Score (WG)} & \textbf{$r$} \\
\midrule
Purpose                 & 662.00 & $<0.001$   & 1.00 & 1.00 & 0.66 \\
Guidelines              & 284.50  & $<0.001$   & 1.00 & 1.00 & 0.84 \\
Limitations             & 610.50 & $<0.001$   & 0.86 & 1.00 & 0.86 \\
Parameter Explanation   & 1104.00 & $<0.001$   & 0.71 & 1.00 & 0.81 \\
Examples                & 54.00  & 0.47      & 1.00 & 1.00 & -0.21 \\
\bottomrule
\end{tabular}
\end{table}

To assess whether source-code grounding improves factual accuracy, we compared augmented descriptions generated without grounding (UG) and with source-code grounding (WG) for MCP-Universe tools whose source code is publicly available in the MCP-Universe benchmark. We define faithfulness as whether the generated component-level claims are supported by the tool source code and input schema, following the prior study by Shahul et al.~\cite{es2024ragas}.

\textbf{Source-code grounding reduces hallucinations and improves the faithfulness of the augmented tool description}. As shown in~\tabref{tab:faithfulness_ug_vs_wg}, source-code grounding improves faithfulness overall, with statistically significant gains for \textit{Limitations}, \textit{Parameter Explanation}, \textit{Purpose}, and \textit{Guidelines}. The strongest improvement appears in \textit{Parameter Explanation}, where the median faithfulness score increases from 0.71 to 1.00. While many tools already achieve high faithfulness without grounding, the Wilcoxon signed-rank test detects improvements because grounding corrects lower-scoring cases, effectively reducing hallucination.

\textbf{However, higher faithfulness alone does not guarantee better agent performance}. We re-executed 92 tasks from Finance, Repository Management, and 3D Design, and found that grounded descriptions solved 26 tasks, while ungrounded descriptions solved 35. McNemar’s exact test reports no statistically significant difference ($p=0.064$). One likely explanation is that many MCP-Universe tools wrap widely used services (e.g., Yahoo Finance, GitHub, Blender APIs) whose common usage patterns are probably present extensively in FM training data. For example, popular open-source training datasets, e.g., ToolRM~\cite{agarwal2025toolrm} and APIGen~\cite{liu2024apigen}, contain Yahoo Finance and GitHub API-related training data. In such cases, source-code grounding improves description faithfulness and reliability, but may not substantially change the model's tool-selection or parameterization decisions during execution. We do, however, observe a complementary effect at the task level: of the 26 tasks solved with grounded descriptions, 5 were not solved with the ungrounded ones, suggesting that grounding can unlock cases where the ungrounded description is insufficient. Identifying which task or tool characteristics make grounding impactful, and how to route between grounded and ungrounded descriptions accordingly, is a promising direction for future work.

\textbf{Therefore, source-code grounding should be viewed as a reliability-oriented safeguard rather than a guaranteed performance optimization}. It is valuable because it reduces hallucinated or unsupported details in augmented descriptions. Its task-level benefits may become more visible for rare, proprietary, or domain-specific tools where the FM cannot rely on prior exposure to infer correct behavior. Future work should therefore study how to combine grounding with conciseness and task-oriented optimization, and whether grounded augmentation yields stronger execution gains for tools outside the FM's training distribution.

\subsection{Impact of Rubric Components in Potential Smell Detection}
\textbf{Rubric-defined smells should be interpreted as potential indicators of risk rather than deterministic predictors of task failure}. OOur rubric is derived from MCP documentation, practitioner guidance, and prompt-engineering taxonomies~\cite{liu2026comprehensive}, so it captures components that are expected to help an FM understand a tool. However, the absence of a tool-description component does not imply that an agent will fail to use the tool properly. This phenomenon is analogous to traditional code smells: just as a code smell in a smaller project might not have as much impact on software maintainability, a missing component for a simple tool may not inherently degrade an agent's performance.

As we observed in RQ2 (\secref{results-rq2}), resolving rubric-defined smells through augmentation improves the task success rate by 5.85 percentage points and the Average Evaluator Score (AE) by 15.12\% across evaluated models. Moreover, performance improved in 54.17\% of the evaluated cases, while 29.16\% remained unchanged, indicating that the rubric components capture information useful for FM-based agents. At the same time, the gains are not universal: 16.67\% of cases regressed, and augmented descriptions increased execution cost. This pattern suggests that the rubric components provide a quality signal, but that their impact also depends on the execution context rather than on component presence alone.

RQ3 further refines this interpretation by showing that not every missing component has the same operational severity. Compact descriptions sometimes match or outperform fully augmented descriptions, especially when the reduced description preserves the tool’s core purpose and the most relevant usage constraints. This divergence is most supported by the ablation of the Examples component. Although examples are traditionally useful in few-shot prompting~\cite{brown2020language}, removing Examples does not significantly degrade performance in our evaluated agent. This does not mean that examples are never useful; rather, it indicates that examples are not uniformly necessary for every MCP tool and may be redundant for simple or familiar tools.

In summary, these findings indicate that, instead of  treating every missing component as equally harmful, future studies should examine when a potential smell is likely to become a performance bottleneck. Relevant factors include tool complexity, model familiarity with the underlying domain, task requirements, and the available context budget. Another important factor is the model's prior exposure to the tool, API, or domain during training. If the model already has sufficient internal knowledge, or if the tool and task are inherently simple, a direct and compact description may be sufficient. Conversely, for unfamiliar domains, complex tasks, or tools with subtle invocation constraints, fully augmented descriptions may be more beneficial because they help bridge the model's comprehension gap.

\section{Implications}\label{sec:implications}
This section outlines the practical and research implications of our results for four key audiences: (i) MCP developers, (ii) ecosystem maintainers (e.g., protocol and registry maintainers), (iii) MCP users (e.g., agent developers), and (iv) researchers. 


\subsection{Implications for MCP Developers}
\textbf{MCP developers should integrate rubric-based smell detection into their review or Continuous Integration (CI) pipelines to prevent the deployment of sub-optimal tool descriptions}. As RQ-1 reveals, 56\% of tools suffer from \textit{Unclear Purpose} and 89.3\% lack \textit{Usage Guidance}, effectively rendering them as stubs rather than functional specifications. These high smell rates indicate that current ad hoc writing practices are insufficient; teams should therefore treat descriptions as first-class engineering artifacts. They need to consider tool description quality as a blocking criterion for release and use automated scanners to detect smells.

\textbf{MCP developers developing new MCP servers or refactoring the existing individual descriptions should prioritize a small set of high-leverage tool description components rather than attempting to fix every aspect simultaneously}. As RQ-1 and Table~\ref{tab:smell_free_combinations} show, 44\% of tools are smell-free on \textit{Purpose} alone, yet this drops to only 2.9\% when all five major components are considered together. On the other hand, RQ-3 demonstrates that compact combinations, such as \textit{Purpose + Guidelines} in Finance, can outperform a tool description that contains all components. Given the practical token limitations of current models, e.g., where large descriptions consume valuable context window capacity of FMs and increase execution cost, developers should first identify and optimize the most impactful components for their tools that convey critical semantic intent with minimal text. Only after establishing these core elements should additional components, such as examples or exhaustive parameter semantics, be introduced selectively, and only in domains where they demonstrably justify the additional token overhead and context window consumption. This token-aware prioritization strategy should help balance quality improvements with resource efficiency, preserving context budget for essential reasoning and reducing unnecessary model invocation costs.

\textbf{MCP developers can consider FM-based \rqtwonoun as a refinement process, but should weigh it against scale, costs, and alternative manual or semi-automated processes}. \tabref{tab:wilcoxon_bo_ao_full} shows that FM-based \rqtwonoun can lift median tool description scores from the 1-2 range to nearly 5.0 across all components. However, blindly relying on FMs can lead to verbose descriptions that unnecessarily consume the context window, especially depending on the scale of the MCP server (hundreds of tools vs. a couple). Moreover, in the RQ-3, we have seen that FMs can sometimes produce confusing instructions in some components, e.g., in the \textit{Limitations} of the \texttt{get\_historical\_stock\_prices} tool. Hence, for developers maintaining servers with few tools, manual refinement may suffice. On the other hand, for larger servers, developers should utilize the semi-automated framework described in~\secref{optimizing-tool-description} to \rqtwoverb the tool description while critically reviewing the output to ensure the \rqtwoactionpast text contains critical operational cues and remains concise. The goal of FM-based \rqtwonoun should be to resolve ambiguity without inflating the token footprint to a point where the agent's token cost cannot justify the operational efficiency anymore.

\subsection{Implications for Ecosystem Maintainers}
\textbf{To enable dynamic context management and reduce token overhead, MCP protocol maintainers should extend MCP specifications beyond the current monolithic description field to structured schema definitions for individual components}. The current MCP specification treats the tool description as a monolithic text blob, obscuring distinct semantic elements such as \textit{Purpose}, \textit{Guidelines}, \textit{Limitations}, \textit{Parameter Explanation}, and \textit{Examples}. Additionally, RQ-3 demonstrates that different subsets of components, e.g., \textit{Purpose + Guidelines} or \textit{Purpose + Guidelines + Limitations + Parameter Explanation}, can preserve core semantic information across different domains, driving equivalent or better success profiles than both baseline and fully \rqtwoactionpast tool descriptions. Hence, by introducing dedicated fields for each component (e.g., distinct JSON fields for Purpose, guidelines, and examples), protocol designers can empower agents to dynamically assemble the most effective description profile at runtime. An agent low on context space could request only the Purpose of tools for initial selection and lazily load guidelines or examples only when a specific tool is invoked, significantly optimizing token window usage. This structural decoupling will allow MCP developers and users both to selectively load, experiment with, or optimize specific components of the description based on the immediate token budget and domain requirements, rather than being forced to consume a fixed text blob.

\textbf{Registry maintainers should integrate rubric-based smell detection and quality scoring as built-in services across MCP registries and marketplaces}. As RQ-1 shows, smells are pervasive across both official and community servers, indicating that no maintainer group consistently produces high-quality descriptions. This finding implies that quality cannot be reliably delegated to individual teams. MCP registries, such as Glama, Smithery, or Cloudflare Workers, existing review processes primarily focus on infrastructure setup or package-level checks, and rarely assess the quality of tool descriptions. Ecosystem maintainers can respond to the pervasive smells by running FM-based scans over published servers, with smell summaries and \rqtwonoun-aware quality badges, and issuing warnings when a tool or server falls below agreed thresholds. Because the FM-based scanner shows stable agreement across models, these diagnostics can function similarly to security advisories in registries such as \texttt{npm}, encouraging developers to submit higher-quality descriptions or maintain multiple \rqtwoactionpast profiles to reduce integration friction.

\subsection{Implications for MCP Users}
\textbf{MCP users should treat tool descriptions as mutable client-side configurations and use them as a cost-effective leverage point, rather than immediately defaulting to larger and more expensive frontier models}. As shown in~\secref{results-rq2}, the smaller-sized \model{Qwen3-Next-80B-A3B-Instruct} model, when equipped with \rqtwoactionpast tool descriptions, achieves performance parity with or even surpasses the significantly larger \model{Qwen3-Coder-480B-A35B} in domains such as \texttt{Finance}, \texttt{Repository Management}, and \texttt{Location Navigation}. This implies that high-quality tool descriptions can serve as an architectural catalyst, enabling the use of smaller, more cost-effective models without sacrificing reliability in specific domains, which MCP users should identify and utilize.

In current practice, however, MCP users typically consume default tool descriptions as fixed artifacts authored and distributed by MCP server developers or vendors, without modification. Our methodology demonstrates that this constraint is not fundamental. By re-purposing the Tool Description Router described in~\secref{mcp-client-calibration}, MCP users can override the default tool description at runtime without modifying server code. This client-side customization will enable MCP users to adapt tool descriptions to their specific domain, model, and context window constraints, providing a practical mechanism for description-level \rqtwonoun until the MCP protocol natively supports equivalent capabilities.

\textbf{MCP users should enforce explicit resource caps, such as maximum step counts, thinking budget, rate limit, or token limits, tailored to the complexity of the domain when deploying \rqtwoactionpast tool descriptions}. As RQ-2 and Table~\ref{tab:overall_metrics_comparison} demonstrate, while \rqtwoactionpast descriptions boost the overall Success Rate (SR) by 5.85 percentage points, they simultaneously inflate the Average Steps (AS) by 67.46\%. This substantial increase in execution overhead implies that richer descriptions improve reasoning at the expense of computational efficiency. Consequently, teams should implement budgeted policies before enabling full \rqtwonoun, allowing higher caps in domains where SR gains justify the overhead (for example Finance or Repository Management) and enforcing lower caps or compact configurations such as \textit{Purpose + Guidelines} in domains where gains are small or negative (for example Web Searching in Table~\ref{tab:optimized_sr_comparison}) to preserve runtime performance.

\subsection{Implications for Researchers}
\textbf{Researchers should investigate holistic mechanisms that improve agent convergence with minimal cost, recognizing that \rqtwoactionpast tool descriptions are one of several levers in a broader efficiency landscape}. RQ-2 and RQ-3 of this study show that \rqtwoactionpast tool descriptions significantly increase task success and evaluator coverage; however, these gains often come with higher step counts or increased token costs. To optimize the flow, emerging approaches such as MCP Zero’s active tool discovery~\cite{fei2025mcp} aim to reduce context size by loading only the tools an agent is likely to need. Similarly, Anthropic has proposed reactive \texttt{tool search}~\cite{anthropic-code-mode} to enable agents to find the correct tools for a task through a semantic search. Moreover, recent code-mode or programmatic tool execution pipelines proposed by Cloudflare~\cite{cloud-flare-code-mode} and Anthropic~\cite{anthropic-code-mode} enable FMs to generate code that orchestrates tool calls directly, thereby avoiding the transport of intermediate tool responses back into the model and reducing both the number of steps and token usage. 

However, the role of description quality within these efficiency mechanisms remains empirically unexplored. It is unknown whether the rubric-\rqtwoactionpast descriptions can improve the recall of dynamic tool discovery of the agents or increase the syntactic correctness and stability of generated orchestration code. If they do, these approaches may achieve the high success rates observed in our experiments without incurring the resource penalties associated with traditional multi-step loops. Researchers should therefore investigate how behaviorally precise, rubric-aligned description variants influence tool search accuracy, static planning quality, and code-generation reliability within these emerging cost-saving architectures.

\textbf{Researchers should extend tool description component ablations to progressive-disclosure mechanisms, e.g., agent skills, and empirically test whether current challenges of these emerging techniques are rooted in under-specified metadata}. Agent skills expose only minimal metadata (e.g., name and description) and load full instructions on demand, aiming to reduce context usage while enabling autonomous invocation. Yet, practitioner reports suggest that models often ignore available skills unless explicitly prompted, indicating a gap between theoretical progressive discovery and observed behavior\footnote{\url{https://scottspence.com/posts/claude-code-skills-dont-auto-activate}, \url{https://www.reddit.com/r/ClaudeAI/comments/1qbc30u/comment/nz9lom3/}}. In this regard, we observed that behaviorally precise components, especially those that encode critical operational cues and explicit usage constraints, can materially shift agent behavior in the RQ-3 results. These observations may suggest that discoverability failures in agent skills may stem from missing or weakly expressed critical cues rather than insufficient metadata volume. Consequently, future research can adopt a similar ablation-style approach on skill metadata to identify the smallest set of high-impact semantic signals that reliably trigger invocation under strict token constraints, preserving the efficiency goals of progressive disclosure while improving its practical effectiveness.

\section{Threats to Validity}\label{sec:threatstovalidity}





\subsection{External validity}
Our dataset comprises 856 tools across 103 MCP servers collected from prior empirical and evaluation studies, and therefore potentially excludes MCP servers that have not yet been evaluated or documented in the literature. As a result, some categories of MCP servers, including proprietary internal deployments or recently introduced servers, may be underrepresented or absent from our dataset. However, the dataset includes a mix of community-maintained open source servers, officially managed open source servers (e.g., GitHub and Playwright), and a small number of officially managed closed MCP servers (e.g., PayPal), covering multiple governance and deployment settings. Moreover, we extract tool descriptions through a dynamic MCP client based reflection mechanism rather than static source code analysis. This design makes the smell detection and \rqtwonoun pipeline agnostic to implementation language and source code availability, and in principle applicable to both open source and closed source MCP servers.

While we detect and optimize smells across 856 tools from 103 MCP servers, performance evaluation is conducted only on the subset of tools and servers included in the MCP-Universe benchmark. Specifically, MCP-Universe covers 202 tools drawn from 18 MCP servers, which represent a strict subset of the full corpus analyzed in RQ1 and RQ2. In addition, due to the high token cost of running the full benchmark, we do not evaluate all models included in MCP-Universe. To mitigate these limitations, we adopt MCP-Universe because it inherently covers a diverse set of tasks spanning multiple domains, MCP servers, and evaluators, enabling robust execution-based assessment without additional sampling. We also intentionally select a heterogeneous model set, covering a frontier proprietary model (GPT-4.1), large open-weight models (\model{Qwen3-Coder-480B-A35B} and \model{GLM-4.5}), and a smaller-sized model (\model{Qwen3-Next-80B-A3B-Instruct}), to capture variation across model families while balancing evaluation cost.

The feasibility of evaluating the fully \rqtwoactionpast tool description depends on the available context window of the models and the verbosity of the tool descriptions. For \model{Qwen3-Next-80B-A3B-Instruct}, we exclude the \textit{Parameter Explanation} and \textit{Examples} components to avoid context overflow. Similarly, for the Browser Automation domain, the excessive length of execution examples necessitates excluding examples for all models. As a result, the evaluation for \model{Qwen3-Next-80B-A3B-Instruct} reflects a partial \rqtwonoun setting, and the full benefits of the proposed pipeline may not be realizable on resource-constrained models or highly verbose tool domains. We justify our choices by prioritizing components that are not otherwise available to the FM. Specifically, while \textit{Parameter Explanation} is omitted, the MCP protocol still provides the input schema to the FM, partially compensating for its absence. In addition, our ablation study in RQ-3 shows that excluding \textit{Examples} has a limited impact on performance in most settings, supporting this design choice under context constraints.

To assess whether MCP-Universe is representative of the broader corpus in terms of quality characteristics, we compared the 202 MCP-Universe tools with the remaining 654 tools. We observed a statistically significant difference in two components: Purpose and Length \& Completeness. But they have only small or negligible effect sizes. Moreover, their median scores are lower for MCP-Universe than for the remaining corpus: 2.00 vs. 2.33 for Purpose and 1.33 vs. 1.67 for Length \& Completeness. Hence, MCP-Universe does not appear to be an unusually high-quality subset of the corpus, and the performance gain observed after augmenting the MCP-Universe tool descriptions did not stem from pre-existing quality differences in these components.

Furthermore, we use the average number of execution steps (AS) as the primary cost proxy, as the MCP-Universe benchmark does not report token-level statistics uniformly across all evaluated models. To mitigate the risk that AS may not reflect token cost, we explicitly measure token usage for \model{Qwen3-Next-80B-A3B-Instruct} across Finance, Repository Management, and 3D Design, where both original and augmented runs are available under the same execution environment. This analysis confirms that token usage increases in the same direction as AS when augmented descriptions are used, supporting AS as a benchmark-compatible proxy for cost in our full analysis. However, the absolute magnitude of token overhead may vary across models and domains not covered by this token-level analysis.

\subsection{Construct validity}
We identified smells by applying a threshold to the component-wise scores of tool descriptions, assuming that scores below the minimum viable threshold indicate meaningful deficiencies. This introduces several construct validity concerns. The choice of score three as the threshold is inherently subjective. Additionally, smell detection further relies on FM-based scoring, which is sensitive to the design of prompts and model-specific preferences. To reduce the subjectivity of the evaluation, we adopt a structured Likert-style analytic scoring over open-ended FM judgments following prior similar studies~\cite{wang2025can}. We justify using FMs rather than human annotators because FMs ultimately interpret tool descriptions within agentic workflows. We further reduce model-specific bias through a multi-model LLM-as-jury setup spanning three model families and report inter-rater reliability. Finally, to validate the jury’s output, we conduct a human evaluation of a representative sample of tools and report agreement between the human evaluators and the jury.

Also, the evaluation and augmentation of tool description components, such as \textit{Purpose}, \textit{Guidelines}, and \textit{Parameter Explanation}, can be optionally grounded in statically extracted source code, but this introduces implementation-specific threats. Source-code grounding is only possible for MCP servers with publicly available codebases and not feasible for proprietary MCP servers in MCP-Universe, such as PayPal. Also, our current extractor supports Python, JavaScript/TypeScript, and Go MCP servers using the common official SDK or FastMCP-style patterns. Tools implemented in other languages, closed-source servers, or unconventional frameworks may be missed, although the approach is extensible. Moreover, many MCP servers wrap upstream APIs, so some limitations may reside outside the local source code. Still, the extracted code, together with input and output schemas, exposes the MCP-level contract and provides stronger grounding than relying only on the original description.

\subsection{Internal validity}
For three models adopted from the MCP-Universe benchmark, the original study does not report per-task SR, AE, or AS, and we do not re-execute these baseline configurations due to the prohibitive computational cost. Consequently, comparisons for these models rely on aggregate baseline metrics reported in prior work, which may differ from our execution environment. In addition, agentic tool-use workflows are inherently non-deterministic, and observed performance differences may partially result from stochastic variation rather than systematic effects of tool description \rqtwoaction. To mitigate this threat, we introduce an additional, smaller and very low-cost model, \model{Qwen3-Next-80B-A3B-Instruct}, for which we execute both baseline runs using the original tool descriptions and \rqtwoactionpast runs under identical conditions. This allows direct before-and-after comparison within the same environment. For this model, we further assess the statistical significance of changes in SR, AE, and AS using appropriate paired statistical tests.

The original MCP-Universe study uses a SERP API-based Google Search MCP server with strict query limits. To ensure the uninterrupted execution of Web Searching tasks, we utilize an alternative Google Search MCP server. Differences in the underlying API platform may impact absolute performance in the Web Searching domain, regardless of the tool's description quality. They may also explain the lack of performance improvement observed across all three MCP-Universe models. To mitigate this issue, we evaluate the Web Searching domain for the newly introduced model, \model{Qwen3-Next-80B-A3B-Instruct}, using the same Google Search MCP server for both the baseline and \rqtwoactionpast tool descriptions. As a result, the statistical analyses for this model control for instrumentation differences, ensuring that measured differences in performance reflect description \rqtwoaction effects, as opposed to search API variability.

The ablation study explicitly selects five domain-model combinations and examines five component configurations for each pair, resulting in a total of 25 runs. All ablation settings retain the \textit{Purpose} component, which we treat as mandatory for correct tool interpretation. Additionally, we do not evaluate all possible permutations of components, as each combination would require a separate benchmark run, which would incur substantial computational costs and time. These design choices may introduce selection bias, as the estimated importance of individual components is derived from subcases already known to be sensitive to \rqtwoaction effects. To mitigate this risk, we include domain-model combinations exhibiting both performance improvements and regressions. We also conduct an ablation that retains all components except \textit{Examples}, indicating that the observed findings are not limited to narrowly selected configurations.
\section{Conclusion}\label{sec:conclusion}
This research provides an extensive empirical characterization of MCP tool descriptions and evaluates the extent to which these descriptions affect FM-based agent execution. By analyzing 856 tools from both official and community-maintained servers and by conducting rubric-guided \rqtwonoun, benchmarking, and controlled studies, we establish tool descriptions as a critical but under-engineered artifact of agentic systems.

Our findings reveal that over 97\% of MCP tools suffer from ecosystem-wide description smells, yet \rqtwoaction these artifacts with all components serves as a powerful architectural lever, enabling smaller-sized open-weight models to achieve performance parity with larger frontier models. Rubric-aligned descriptions with all components enhance agent performance, resulting in a 15.12\% increase in the Average Evaluator Score and a 5.85 percentage point improvement in task success rates. However, these gains require more execution steps and higher costs, and are not universal across all domain-model combinations, highlighting the need for cost-aware and context-sensitive \rqtwonoun. On that front, our ablation study shows that compact combinations of high-impact components can achieve behavioral alignment comparable to fully \rqtwoactionpast descriptions while reducing cost overhead. 

Collectively, these findings call for a shift toward treating tool descriptions as configurable engineering artifacts, motivating structured, component-aware protocol designs that support dynamic, cost-aware context management in MCP-enabled agents. Future work should investigate how description quality impacts emerging cost-reduction techniques, such as dynamic tool search and code-mode execution of MCP, and whether rubric-aligned descriptions can enhance tool retrieval and orchestration, enabling future agents to achieve high reliability without incurring the cost penalties of traditional multi-step loops.



\section*{Disclaimer}
We employed ChatGPT-5.2, Anthropic Opus-4.5, and Gemini-3 Pro-Preview solely for copy-editing and table layout preparation, consistent with IEEE and ACM policies governing AI use in scholarly work.

\bibliographystyle{ACM-Reference-Format}
\bibliography{main.bib}
\appendix
\section{Appendix}\label{sec:appendix}

\subsection{Prompts used by the LLM-Jury}\label{prompt-jury}
\subsubsection*{Judge Tool Description Quality Using a Six-Component Rubric}

You are grading a tool description. Score each component from 1 to 5, then provide an overall quality score (0--100), a justification, and improvement recommendations.

\paragraph{Scoring Rubric (1--5 scale for each component)}

\begin{enumerate}
    \item \textbf{Purpose (What the tool does)}
    \begin{itemize}
        \item \textbf{5/5}: Clearly explains function, behavior, and return data with precise language.
        \item \textbf{4/5}: Explains function and behavior with minor ambiguity.
        \item \textbf{3/5}: Basic explanation present but lacks behavioral details.
        \item \textbf{2/5}: Vague or incomplete purpose statement.
        \item \textbf{1/5}: Purpose unclear or missing.
    \end{itemize}

    \item \textbf{Usage Guidelines (When to use or not use)}
    \begin{itemize}
        \item \textbf{5/5}: Explicitly states appropriate use cases and when not to use; includes disambiguation if the tool name is ambiguous.
        \item \textbf{4/5}: States when to use with minimal guidance on when not to use.
        \item \textbf{3/5}: Implies usage context but lacks explicit boundaries.
        \item \textbf{2/5}: Usage context unclear or overly generic.
        \item \textbf{1/5}: No usage guidance provided.
    \end{itemize}

    \item \textbf{Limitation (Caveats and boundaries)}
    \begin{itemize}
        \item \textbf{5/5}: Clearly states what the tool does not return, scope boundaries, and important constraints.
        \item \textbf{4/5}: Mentions main limitations but misses some edge cases.
        \item \textbf{3/5}: Vague or incomplete limitation statements.
        \item \textbf{2/5}: Minimal or implied limitations only.
        \item \textbf{1/5}: No limitations or caveats mentioned.
    \end{itemize}

    \item \textbf{Parameter Explanation (Input clarity)}
    \begin{itemize}
        \item \textbf{5/5}: Every parameter is explained with type, meaning, behavioral effect, and required or default status.
        \item \textbf{4/5}: Most parameters are explained with minor omissions.
        \item \textbf{3/5}: Basic parameter information is present but lacks behavioral impact.
        \item \textbf{2/5}: Parameters are listed without meaningful explanation.
        \item \textbf{1/5}: Parameters are not explained or only provided in schema form.
    \end{itemize}

    \item \textbf{Examples vs. Description Balance}
    \begin{itemize}
        \item \textbf{5/5}: Description is self-sufficient; examples, if any, supplement rather than replace the explanation.
        \item \textbf{4/5}: Mostly descriptive with minor reliance on examples.
        \item \textbf{3/5}: Even mix of description and examples.
        \item \textbf{2/5}: Over-relies on examples with minimal prose.
        \item \textbf{1/5}: Only examples are provided with no descriptive explanation.
    \end{itemize}

    \item \textbf{Length and Completeness}
    \begin{itemize}
        \item \textbf{5/5}: Four or more sentences of substantive, well-structured prose covering all aspects.
        \item \textbf{4/5}: Three to four sentences with good coverage.
        \item \textbf{3/5}: Two to three sentences that are somewhat complete.
        \item \textbf{2/5}: One to two sentences that are too brief.
        \item \textbf{1/5}: Single phrase or fragment.
    \end{itemize}
\end{enumerate}

\paragraph{Input}

\begin{verbatim}
{tool_payload}
\end{verbatim}

\paragraph{Output Format (JSON)}

\begin{verbatim}
{
  "scores": {
    "purpose": 1-5,
    "usage_guideline": 1-5,
    "limitation": 1-5,
    "parameter_explanation": 1-5,
    "examples_balance": 1-5,
    "length_completeness": 1-5
  }
  "label": "Good" | "Bad",
  "reason": "One sentence justification",
  "improvement_needed": [
    "comma separated list of specific weak areas with scores <= 3"
  ]
}
\end{verbatim}

\textbf{Labeling rules:}

A description is labeled \textbf{Bad} if:
\begin{itemize}
    \item Any of the six rubric dimensions score below 3, or
    \item Examples replace the description instead of supporting it.
\end{itemize}

A description is labeled \textbf{Good} only if:
\begin{itemize}
    \item All six dimensions score 3 or higher, and
    \item All requirements in components 1 through 6 are satisfied.
\end{itemize}

\subsection{Prompts used by Tool Description Augmentor}\label{augmentor-prompt}
\subsubsection{Task Generation Prompt}

\begin{verbatim}
_SYSTEM_PROMPT = """\
You are an expert benchmark designer for AI agent evaluation.
Your job is to create realistic, diverse benchmark tasks that test an AI agent's \
ability to use a specific MCP (Model Context Protocol) tool correctly.

Each task must be a self-contained natural-language question that:
- Has a single, unambiguous correct answer obtainable by calling the tool once.
- Is written in plain, conversational English exactly as a real user would type \
  it — short and direct, not formal or verbose.
  Good: "What was AAPL's closing price on 2 Jan 2024?"
  Bad:  "Please provide the historical stock price for the ticker symbol AAPL on \
  January 2nd, 2024, using the get_historical_stock_prices tool."
- Specifies all input values needed (dates, tickers, names, units, etc.) so the \
  agent does not need to guess.
- Uses varied, realistic values across tasks — avoid repeating the same ticker, \
  date, name, or parameter value in multiple tasks for the same tool.

Difficulty levels:
  medium — standard, realistic usage of the tool with normal inputs.
  hard   — targets an edge case, corner case, near-limit input, boundary value, \
            unusual encoding, or a parameter combination that exercises the tool's \
            limits or error handling.
  very hard — significantly more demanding than hard; should combine multiple \
            constraints, ambiguity-resolution requirements, brittle edge cases, \
            or tool-specific failure boundaries while still remaining solvable \
            with a single tool call.

Return ONLY a JSON object in this exact schema — no prose, no markdown fences:
{
  "tasks": [
    {
      "difficulty": "medium" | "hard" | "very hard",
      "question": "<concise, user-style question>",
      "correct_answer": "<the exact answer the tool should return>",
      "rationale": "<one sentence explaining why this difficulty was chosen>"
    }
  ]
}
"""
\end{verbatim}

\subsubsection{Jury LLM Prompt}
\begin{verbatim}
_SYNTHESIS_SYSTEM = """\
You are a technical writer creating documentation for an MCP (Model Context Protocol) tool.

You will be given a set of observed tool executions: each entry shows the exact
arguments passed to the tool and the exact response it returned (or the error
it raised).  Your job is to turn this evidence into a concise, factual
``additional_description`` block that helps an AI agent use the tool correctly.

The block must contain two clearly labelled sections:

**Examples**
  One numbered example per distinct observation (do not invent scenarios that
  are not present in the data).  Each example must:
  - State its purpose in one sentence.
  - Show the exact JSON arguments in a code block.
  - Show the exact response (or a faithful excerpt  of maximum 20 lines) in a code block.
  - Prefer observations that demonstrate an edge case, error condition, or
    boundary value over plain happy-path calls.

**Limitations**
  A bullet-point list of limitations, constraints, and gotchas distilled
  strictly from the observations: required parameter formats, valid value
  ranges, known error conditions, surprising behaviours (e.g. isError=false
  even on error), and anything an agent should watch out for.
  Every bullet must be backed by something actually seen in the observations.
  Do NOT copy generic API-design advice; only include tool-specific findings.

Write in plain, direct English.  Do not invent facts not supported by the
observations.  Do not repeat the tool description verbatim.
Return ONLY the documentation text — no JSON wrapper, no markdown heading for
the tool name itself.
"""
\end{verbatim}

%
%
%
%
%
%
%
%

\end{document}